\documentclass[sigconf]{acmart} 

\AtBeginDocument{%
  }

\setcopyright{acmlicensed}
\copyrightyear{2025}
\acmYear{2025}
\setcopyright{cc}
\setcctype{by}
\acmConference[CHI '25]{CHI Conference on Human Factors in Computing Systems}{April 26-May 1, 2025}{Yokohama, Japan}
\acmBooktitle{CHI Conference on Human Factors in Computing Systems (CHI '25), April 26-May 1, 2025, Yokohama, Japan}\acmDOI{10.1145/3706598.3713788}
\acmISBN{979-8-4007-1394-1/25/04}

\newcounter{answernum}
\newcommand{\newanswer}[1][]{\refstepcounter{answernum}{#1}\textbf{C{\theanswernum}}:}

\def\eg{{\it e.g.}}

\def\ie{{\it i.e.}}
\def\etal{{\it et al.}}

\newcommand{\red}{\textcolor[rgb]{0,0,0}}

\newcommand{\rrred}{\textcolor[rgb]{0,0,0}}

\newcommand{\rred}{\textcolor[rgb]{0.757,0.153,0.212}}

\newcommand{\green}{\textcolor[rgb]{0.180, 0.518, 0.349}}
\definecolor{lightgray}{rgb}{0.89, 0.89, 0.89}
\usepackage{newunicodechar}
\usepackage[utf8]{inputenc}
\usepackage[T1]{fontenc}    
\usepackage{pifont}
\usepackage{newunicodechar}
\newunicodechar{✓}{\green{\ding{51}}}
\newunicodechar{✗}{\rred{\ding{55}}}

\usepackage{xcolor}
\usepackage{listings}
\usepackage{multirow}
\usepackage{booktabs}
\begin{document}

\title{WanderGuide: Indoor Map-less Robotic Guide\\for Exploration by Blind People}

\author{Masaki Kuribayashi}
\affiliation{
  \institution{Waseda University}
  \city{}
  \country{}}
\affiliation{
   \institution{Miraikan - The National Museum of Emerging Science and Innovation}
   \city{Tokyo}
\country{Japan}}

\author{Kohei Uehara}
\affiliation{
   \institution{Miraikan - The National Museum of Emerging Science and Innovation}
   \city{Tokyo}
\country{Japan}}

\author{Allan Wang}
\affiliation{
   \institution{Miraikan - The National Museum of Emerging Science and Innovation}
   \city{Tokyo}
\country{Japan}}

\author{Shigeo Morishima}
\affiliation{
  \institution{Waseda Research Institute for Science and Engineering}
  \city{Tokyo}
  \country{Japan}}

\author{Chieko Asakawa}
\affiliation{
   \institution{Miraikan - The National Museum of Emerging Science and Innovation}
   \city{Tokyo}
   \country{Japan}}

\renewcommand{\shortauthors}{Kuribayashi et al.}

\begin{abstract}
Blind people have limited opportunities to explore an environment based on their interests. While existing navigation systems could provide them with surrounding information while navigating, they have limited scalability as they require preparing prebuilt maps. Thus, to develop a map-less robot that assists blind people in exploring, we first conducted a study with ten blind participants at a shopping mall and science museum to investigate the requirements of the system, which revealed the need for three levels of detail to describe the surroundings based on users' preferences. Then, we developed WanderGuide, with functionalities that allow users to adjust the level of detail in descriptions and verbally interact with the system to ask questions about the environment or to go to points of interest. The study with five blind participants revealed that WanderGuide could provide blind people with the enjoyable experience of wandering around without a specific destination in their minds.
\end{abstract}


\begin{teaserfigure}
  \includegraphics[width=\textwidth]{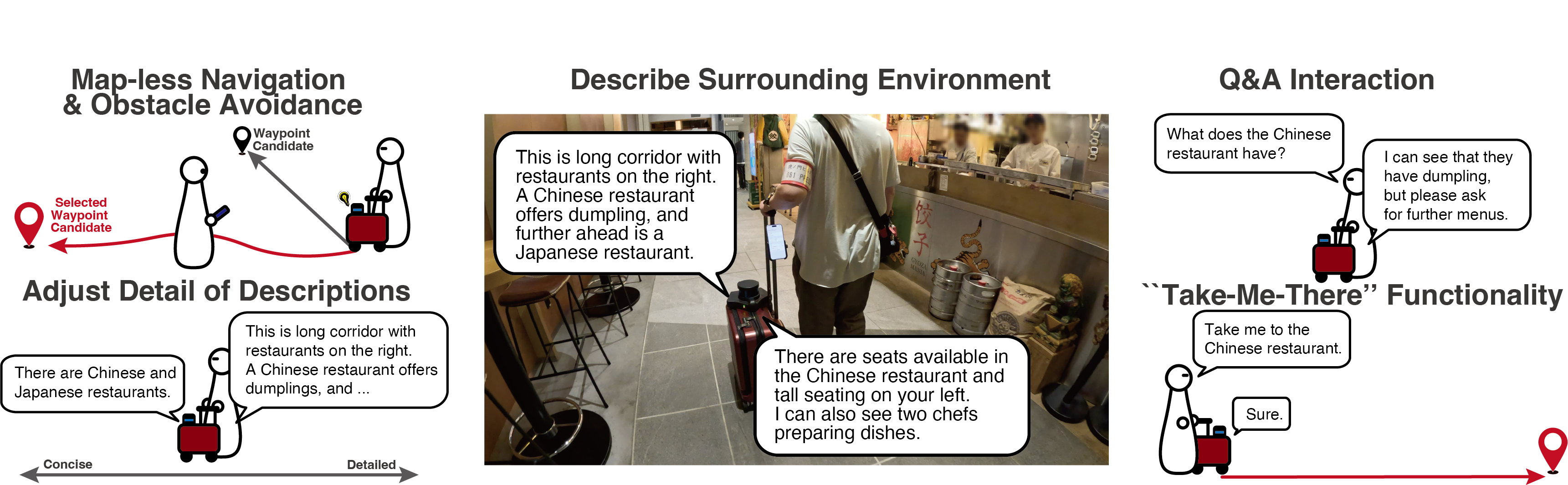}
  \caption{Five core functionalities of WanderGuide. The system assists users in recreational exploration by explaining the surrounding environment through images obtained from the robot's camera. Users can adjust the level of detail and ask questions about their surroundings. Additionally, the system can guide users to locations they have visited before.}
  \Description{
    The figure shows the five core functionalities of WanderGuide. These functionalities include: Map-less Navigation and Obstacle Avoidance: WanderGuide selects potential waypoint candidates using visual information from the environment. It helps users navigate around obstacles and choose a path to their destination without depending on a pre-existing map. The figure illustrates the robot predicting several possible waypoints, selecting one, and navigating while avoiding a person. Describe Surrounding Environment: WanderGuide communicates details about the environment's layout. In the figure, the system is describing, "This is a long corridor with restaurants on the right. A Chinese restaurant offers dumplings, and further ahead is a Japanese restaurant. There are seats available in the Chinese restaurant, and tall seating on your left. I can also see two chefs preparing dishes." Adjust Detail of Descriptions: Users can adjust the level of detail in descriptions. A concise description might say, "There are Chinese and Japanese restaurants," while a more detailed version could include, "This is a long corridor with restaurants on the right. A Chinese restaurant offers dumplings, and ..." Question and Answering Interaction: WanderGuide allows users to ask questions about their surroundings. In the figure, the user is asking, "What does the Chinese restaurant have?" and WanderGuide could respond, "I can see they have dumplings, but please ask for further menus." "Take-Me-There" Functionality: WanderGuide allows users to request to be guided to a specific location. In the figure, the user is saying, "Take me to the Chinese restaurant," and WanderGuide guides the user to that location. 
  }
  \label{fig:teaser}
\end{teaserfigure}


\maketitle

\begin{table*}[]
\caption{Comparison to previous work. Our work explores a unique characteristic that has not been investigated in the past. \rrred{In the Purpose row, ``Navigation'' refers to systems primarily designed to guide users to their intended destinations, while ``Perception'' refers to those focused on understanding the surrounding environment. ``Multi-purpose'' refers to systems capable of performing various tasks, and ``Exploration'' refers to those designed for navigating and enjoying facilities, characterized by constantly discovering and changing goals (\textit{e.g.}, window-shopping~\cite{kamikubo2024we,Kaniwa2024ChitChatGuide}).}}
\label{tab:relatedwork}
\resizebox{\textwidth}{!}{%
\begin{tabular}{lcccc}
\toprule
  System &
  Map-less &
  Purpose &
  \begin{tabular}[c]{@{}c@{}}Independent from\\
  Human Assistance\end{tabular} &
  Device \\
  \midrule
NavCog~\cite{sato2019navcog3}           & ✗ & Navigation  & ✓ & Smartphone \\
CaBot~\cite{guerreiro2019cabot}            & ✗ & Navigation  & ✓ & \textbf{Robot}      \\
Tactile Compass~\cite{liu2021tactile} & ✗ & Navigation & ✓ & Handheld Device \\
ChitChatGuide~\cite{Kaniwa2024ChitChatGuide}    & ✗ & \textbf{Exploration} & ✓ & Smartphone \\
Kayukawa \etal~\cite{kayukawa2023enhancing}         & ✗ & \textbf{Exploration} & ✗ & \textbf{Robot} \\
Corridor-Walker~\cite{Kuribayashi2022CorridorWalker}  & ✓ & Navigation  & ✓ & Smartphone \\
Snap\&Nav~\cite{Kubota2024Snap}        & ✓ & Navigation  & ✗ & Smartphone \\
PathFinder~\cite{kuribayashi2023pathfinder} & ✓ & Navigation  & ✗ & \textbf{Robot} \\
WorldScribe~\cite{chang2024worldscribe}    & ✓ & Perception & ✓ & Smartphone \\
GPT-4o Demo~\cite{GPT4o}           & ✓ & Perception     & ✓ & Smartphone \\
MLLM Powered Applications (\eg, Seeing AI~\cite{SeeingAI})          & ✓ & Perception     & ✓ & Smartphone \\
RSA~\cite{BeMyEyes,Aira} & ✓ & Multi-Purpose     & ✗ & Smartphone \\
\midrule
WanderGuide      & ✓ & \textbf{Exploration} & ✓ & \textbf{Robot} \\
\bottomrule
\end{tabular}
}
\end{table*}

\section{Introduction}
Exploration is a fundamental skill that allows one to gain familiarity with novel environments that blind people do not know. 
Sighted people explore by visually perceiving points of interest (POI) and navigating to desirable destinations. 
However, blind people face significant challenges in independently exploring new environments~\cite{Engel2020travelling,muller2022traveling}. 
They typically rely on sighted assistants, such as friends or family members, to help them navigate and describe their surroundings.
Unfortunately, these assistants are not always readily available, resulting in limited opportunities for blind people to explore independently.

\red{
Over the past recent years, various guide systems~\cite{bineeth2020blindsurvey, sulaiman2021analysis, manjari2020survey}, that are aimed for navigation~\cite{sato2019navcog3,li2016isana} or exploration~\cite{Kaniwa2024ChitChatGuide,kayukawa2023enhancing}}, have been developed to guide blind people and provide details about surrounding POIs in the environment. 
These systems typically rely on prebuilt maps and localization infrastructure 
\red{(\eg, Bluetooth Low Energy (BLE) beacons~\cite{sato2019navcog3,murata2018smartphone,kim2016navigating,chen2015blindnavi,InclusiveNavi} and ultrawide-bandwidth beacons~\cite{lu2021assistive})}
that are highly customized to the environments to continually update their current locations and offer turn-by-turn navigation guidance.
\red{
As access to the prebuilt maps also allows these systems to convey information about nearby POIs while navigating, some systems are specialized in assisting exploration activity~\cite{Kaniwa2024ChitChatGuide,kayukawa2023enhancing}.
}
However, only a limited number of \red{guide systems (\eg, InclusiveNavi~\cite{InclusiveNavi} and BlindSquare~\cite{BlindSquare})} are publicly \red{deployed} because configuring and maintaining prebuilt maps and localization infrastructure is expensive, and it is infeasible for them to be deployed in unseen environments. 
\red{
Several systems that do not require maps, as well as remote sighted assistance (RSA)~\cite{kamikubo2020support,Aira,BeMyEyes}, have been developed to guide blind people in various locations~\cite{kuribayashi2023pathfinder,Kuribayashi2022CorridorWalker,fallah2012user,lacey2000context}. 
However, these systems primarily focus on navigation to target destinations, not exploration, thus providing only navigation-related information to users (\eg,  intersections~\cite{Kuribayashi2022CorridorWalker,kuribayashi2023pathfinder} and signs~\cite{kuribayashi2023pathfinder}). 
These systems are also not independent from human assistance.
}
To promote social inclusion and equality for blind people, there is a need to develop a \textit{map-less} guide system that assists blind people in exploring diverse novel locations without relying on prebuilt maps or infrastructure.

\red{
To bridge the gaps and address the shortcomings of existing systems, we developed a system with the following characteristics as shown in Table~\ref{tab:relatedwork}: 1. Our system does not rely on prebuilt maps or preinstalled infrastructure. 2. Our system focuses on exploration. 3. Our system does not require supplementary assistance from humans. 4. Our system can automatically guide users physically during exploration. None of the prior systems possess the combination of all these characteristics.
Given that the design space for a map-less exploration guide robot remains underexplored, this work aims to investigate and establish the key components of such a system.
We begin by selecting a wheeled robot platform as the device. 
The decision to use a wheeled robot is based on its ability to autonomously guide blind users. 
It alleviates the challenge of navigation, which is cognitively demanding while learning about the surrounding environment.
Additionally, we equip the robot system with the ability to convey real-time information about the surrounding environment to users using natural language, accomplished through a multimodal large language model (MLLM~\cite{GPT4o}). 
}

Using our prototype system, we employed an iterative process with the direct involvement of target users to develop our system. 
In the formative study, the participants were asked to follow the robot, which was controlled in a Wizard-of-Oz fashion~\cite{riek2012wizard-of-oz}, along predetermined routes while listening to the environment descriptions.
The study revealed three groups of user preferences in the system's descriptions with respect to varying levels of details in the descriptive information received.
It also revealed requirements in certain functionalities, such as revisiting locations where the system had mentioned, specifying directions to proceed, and obtaining in-depth information through question-and-answering (Q\&A) functionality.

In the second stage, taking the lessons learned from the first study, we present \textit{WanderGuide}, a map-less exploration system for blind people (Fig.~\ref{fig:teaser}).
Taking into consideration the previously discovered three groups of user preferences, the system offers three modes for describing the surroundings: (1) Detailed description --- in-depth information with high granularity, (2) Balanced-Length description --- balanced level of information, and (3) Concise description --- minimal but essential details for obtaining quick awareness. 
We also implemented various new features based on the feedback received from the first study, which includes adopting a high-resolution fisheye camera for better perception of the surrounding environment, allowing users to verbally interact to query about the environment and set explored POIs to be navigation destinations, and allowing users to use directional buttons to control the robot for navigation towards the direction of interest. 
Our system is also fully integrated with the automatic mapping, localization, map-less navigation, and obstacle avoidance functions of the wheeled mobile robot.

Finally, we conducted a main user study with five blind participants, who were asked to freely explore two floors of the science museum.
All participants appreciated the experience of wandering freely without a fixed destination, and they expressed their desire to use the system to explore both familiar and unfamiliar areas. 
Participants also highlighted the need to incorporate recognition of auditory cues from the environment.
Additionally, differences in how they interacted with the system were observed: one frequently used buttons to guide the robot towards their areas of interest, one passively followed the robot, and others often asked questions. 
We also identified a limitation in the system's MLLM when conveying detailed information about the surroundings, such as identifying specific names of objects, which suggests the need for further development in how we input information into the MLLM for exploration purposes.

\red{
To the best of our knowledge, our work is the first to investigate the design space of a map-less system for blind people to explore independently.
To this end, we made the following contributions.}
\begin{itemize}
    \item \red{We formulated the requirements for the system through a formative study, such as the ability to adjust the level of description based on user preferences and to guide users to previously visited locations of interest, thereby enhancing the exploration experience.}
    \item \red{
     We developed a full stack map-less exploration system that consists of a waypoint detection algorithm and an MLLM-based perception interaction system on top of an existing navigation guide robot. Additionally, we integrated several functionalities based on the formative study that facilitates the exploration experience.
    }
    \item \red{We confirmed key design requirements, such as varying the level of descriptions based on user preferences through a usability study. We also gained further insights into users' interaction preferences and into design implications for improving the system, including better recognition of audio cues.}
\end{itemize}

\rrred{
The codes of the system are publicly available in the following link: \url{https://github.com/chestnutforestlabo/WanderGuide}.
}
\section{Related Work}
\subsection{Exploration for Blind People}
Previous research has emphasized the importance of exploration for blind people to familiarize themselves with the environment~\cite{jain2023want} or for enjoying recreational areas where exploration is essential (\eg, museums~\cite{kayukawa2023enhancing} or shopping malls~\cite{kamikubo2024we}).
The investigation by Engel~\etal~\cite{Engel2020travelling} shows that 59.4\% of the blind population in the study travels to unfamiliar buildings several times a week, but often cannot explore independently, because they rely on sighted assistants with limited availability.
While learning routes and POIs in a building could also be achieved by searching online~\cite{Engel2020travelling}, using interactive maps~\cite{wang2022bentomuseum,nagassa20233d,sargsyan20233d,poppinga2011touchover} or applications~\cite{india2021vstroll,guerreiro2017virtual}, on-site exploration by blind people is also important because by doing this, they receive rich sensory information and gain a better sense of independence~\cite{kamikubo2024we}.
This overall experience motivates them to explore by themselves~\cite{guerreiro2019airport,kamikubo2024we}. 
Focusing on on-site exploration, researchers have investigated the information needs of blind people~\cite{hoogsteen2022beyond,williams2014just,banovic2013uncovering,jain2023want,kamikubo2024we}, and found that it is essential to include high-level understanding of the environment (\eg, layout information~\cite{jain2023want}) as well as specific details such as the names of the shops and brands~\cite{banovic2013uncovering}.
Additionally, researchers noted that safety during navigation is crucial, as safety concerns can dominate the cognitive load and impede the rich exploration experience~\cite{cai2024navigating,zhang2023follower,jain2023want}. 

\subsection{Assistance Systems for Blind People To Explore}
\red{
Robotic guide systems have the advantage of addressing the mobility challenges of blind people with their automatic guidance capability.
CaBot~\cite{guerreiro2019cabot}, the first guidance robot that adopted the form of a suitcase, guides users to specified destinations while referring to prebuilt maps or using an object detector to convey surrounding information.
Among them, some are specialized in exploration~\cite{kayukawa2023enhancing,asakawa2018present,asakawa2019independent}.
A robot system by Kayukawa~\etal~\cite{kayukawa2023enhancing} allows users to explore by interactively setting destinations on a smartphone and by calling a museum guide to explain the surroundings.
However, both systems heavily rely on prebuilt maps and operate in limited locations where the destinations are readily available.
Ultimately, our goal is to develop a system that does not require prebuilt maps and enables blind people to explore independently, \ie, without relying on staff assistance within the facility.
}

Navigation systems for blind people that do not rely on prebuilt maps and infrastructure, \ie, \textit{map-less navigation systems}, have also been proposed in prior research. 
Besides real-time perception outcomes, these systems primarily depend on externally sourced route information, such as prior route knowledge from blind users~\cite{lacey2000context,Kuribayashi2022CorridorWalker}, routes described by nearby pedestrians~\cite{ranganeni2023exploring,kim2023transforming,kuribayashi2023pathfinder}, and analyzed images of floor maps captured in buildings~\cite{Kubota2024Snap}.
\red{
For example, PathFinder~\cite{kuribayashi2023pathfinder} is a map-less navigation robot system designed to guide blind users to their destinations based on predefined routes. 
The system autonomously navigates users by utilizing an intersection detection algorithm~\cite{yang2021graph} and a sign recognition algorithm~\cite{kuribayashi2023pathfinder}. 
These algorithms enable users to determine the correct direction to proceed at key decision points. 
The system's evaluation found that it is necessary to include functionality that takes users back to their starting location after reaching their destination when navigating unfamiliar buildings.
}
However, these map-less navigation systems are tasked with reaching a specific destination and are not suitable for exploration, as they only focus on providing information related to reaching the destination (\eg, intersections and signs~\cite{kuribayashi2023pathfinder}). 
\red{
In an exploration scenario, any information about the environment may prove valuable, such as layout information~\cite{jain2023want}. 
In our study, we aim to explore the underexplored design space of map-less guide robots for exploration purposes, such as how the system should describe surroundings and what are the task-specific functionality requirements.
}

\subsection{Autonomy and Control Methods of Assistant Systems}
Researchers have emphasized autonomy, \ie, the ability for users to select destinations and routes according to their interests, as an important factor for exploratory activities~\cite{Kaniwa2024ChitChatGuide,kayukawa2022HowUsers}. 
To this end, researchers have investigated various control methods based on user inputs~\cite{ranganeni2023exploring,zhang2023follower}. 
For example, systems with prebuilt maps adopted selecting destinations from a premade list of stores~\cite{sato2019navcog3,kayukawa2022HowUsers}, or via conversation~\cite{sato2019navcog3,Kaniwa2024ChitChatGuide}. 
In the case of map-less systems, researchers have adopted feedback-based closed-loop processes to leverage both human inputs and system control. 
Examples include users specifying proceeding directions at intersections while the robot provides automated guidance to the next intersections~\cite{kuribayashi2023pathfinder,jain2023want,ranganeni2023exploring,lacey2000context,hwang2022system}. 
Zhang~\etal~\cite{zhang2023follower} reported that the preferred level of control by blind people may vary depending on context.
Therefore, we examine the level of control between users and robots under the novel exploration context. 

\subsection{Scene Description for Blind People}
\red{Knowledge of surrounding information is crucial for blind people to explore~\cite{Kaniwa2024ChitChatGuide}.} Tools for blind people to understand their surrounding environment have been commercially deployed and the topic remains an ongoing area of research.
\red{
While researchers have proposed tools using visual captioning~\cite{saha2019closing} and question-answering models~\cite{gurari2018vizwiz} to help blind users understand scenes, these often fail to provide accurate descriptions at diverse locations~\cite{delloul2022image}.
Alternatively, RSA applications (\eg, Aira~\cite{Aira} and BeMyEyes~\cite{BeMyEyes}) have long been a practical aid for providing blind people with surrounding scenes.
However, RSA systems are not suitable for our task, as the service quality depends heavily on the sighted assistance provided~\cite{kamikubo2020support}. RSA services may also not be sustainable for extended use until users feel fully satisfied with their exploration experience.
}
With the emergence of LLMs and MLLMs, scene describing systems (\eg, Seeing AI~\cite{SeeingAI}, BeMyAI~\cite{BeMyAI} and GPT4o-demo~\cite{GPT4o}) have been developed, which enable blind people to understand scenes in diverse scenarios~\cite{gonzalez2024investigating,xie2024emerging}. 
\red{
ChitChatGuide~\cite{Kaniwa2024ChitChatGuide} employs LLMs to interpret predetermined maps and deliver exploration-related information during navigation to a specified destination. 
However, unlike our system, it relies on prebuilt maps and lacks the capability to provide real-time information.
MLLM-based systems, such as WorldScribe~\cite{chang2024worldscribe}, offer real-time information by analyzing captured images. 
WorldScribe~\cite{chang2024worldscribe} also adapts the level of description based on user context, such as the speed at which the device is moved.
Our WanderGuide similarly provides three levels of descriptions but the selection is adjusted based on individual user preferences rather than situational context.
On the system level, the core distinction is that WanderGuide combines MLLM with a navigation robot, allowing users to concentrate fully on the descriptions of novel environments and navigate to interested places. This combination makes WanderGuide particularly well-suited for \textit{exploration while navigating}.}
\section{System Design Focus}
\label{sec:system_design}
Our goal is to finalize a system that assists blind people in exploring an indoor environment independently.
In this section, we describe the key design elements of the system.

\begin{figure*}
    \centering
    \includegraphics[width=1\linewidth]{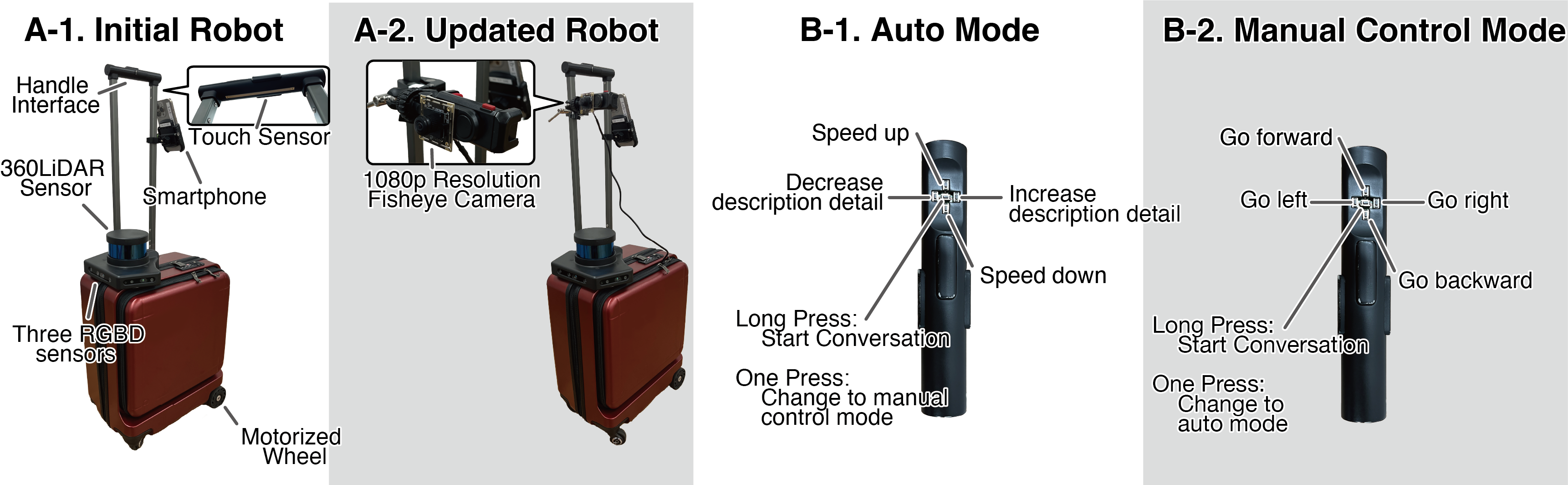}
    \caption{Image of the robot and handle interface used in the study. Panel A-1 shows the robot used in the formative study, while Panel A-2 presents the robot used in the main study. Panels B-1 and B-2 illustrate the mapping of the handle interface buttons' functions, depending on the selected navigation mode.}
    \label{fig:device&ui}
    \Description{The image consists of four panels labeled A-1, A-2, B-1, and B-2. Panels A-1 and A-2 display a suitcase-shaped device used in the robot, while Panels B-1 and B-2 illustrate the button mapping of the handle interface for navigation. Panel A-1 depicts the initial robot used in the formative study. The robot is red and resembles a suitcase. The panel outlines six key components: The handle interface is located where a typical suitcase handle would be and includes five buttons. A smartphone is mounted on the back of the handle using a mounting device. A touch sensor, positioned under the handle, detects when users are touching it. Three RGBD sensors, situated on the front of the robot, are used for depth and color sensing to assist in obstacle detection and navigation. A 360-degree LiDAR sensor is mounted on top of the robot, on top of the three cameras. Motorized wheels at the base provide mobility, allowing the robot to move autonomously or in response to user input. Panel A-2 shows the updated robot used in the main study. The most significant change is the addition of a 1080p-resolution fisheye camera, positioned near the handle to capture images from a higher point of view. Panel C-1 illustrates the button mapping when the robot is in Automated Navigation Mode: The up button increases speed, and the right button decreases it. The left and right buttons adjust the level of detail. A long press on the middle button triggers conversation mode, while a single press switches to manual control mode. Panel B-2 displays the button mapping when the robot is in Manual Control Mode: The up button moves the robot forward, the back button moves it backward, the left button turns it left, and the right button turns it right. A long press on the middle button triggers conversation mode, while a single press returns the robot to automated navigation mode.
    }
\end{figure*}

\subsection{Device}
Assistance systems for blind people have been proposed in various devices, such as smartphones~\cite{presti2019watchout}, \red{handheld haptic devices~\cite{spiers2016outdoor,choiniere2016development,liu2021tactile},} wearable devices~\cite{li2016isana}, cane-like devices~\cite{ranganeni2023exploring} and robots~\cite{liu2024dragon}.
Each type of device offers unique advantages - Smartphones \red{and handheld haptic devices are portable; Smartphones} are also widely used by blind people~\cite{morris2014blind,martiniello2022exploring}; Wearable devices free the user's hands~\cite{lee2014wearable}; Cane-like devices resemble traditional canes~\cite{ranganeni2023exploring}; And robots are able to autonomously guide users~\cite{guerreiro2019cabot}.
\red{While handheld devices~\cite{presti2019watchout,spiers2016outdoor,choiniere2016development,liu2021tactile,ranganeni2023exploring} have often been used due to their portability, in exploration scenarios, they require users to point the devices in their directions of interest while navigating around unfamiliar locations and obstacles, which involve high cognitive load.
Thus}, we chose robots because of their autonomous navigation and obstacle avoidance capabilities. 
This allows users to concentrate on learning the environment~\cite{cai2024navigating,zhang2023follower,jain2023want}. 
In particular, we adopted a wheeled robot~\cite{guerreiro2019cabot,zhang2023follower,wang2022can}.
While wheeled robots are unable to navigate stairs like quadruped robots~\cite{cai2024navigating}, blind users often find wheeled robots more suitable due to their silence and stability~\cite{wang2022can}.
Our assumption is that the devices should ensure the users' safety during navigation and allow users to focus on exploration. 
As a result, the findings in our study can be extended to any similar devices other than wheeled robots. 

\subsection{Describing Scenes}
Previous navigation systems relied on hardcoded information~\cite{sato2019navcog3,Kaniwa2024ChitChatGuide} or simple image captioning models~\cite{saha2019closing} to provide scene descriptions. 
They only convey information related to navigating to destinations. 
In exploratory tasks, any information and details could be relevant. 
Therefore, we decided to use MLLM, a foundational model capable of recognizing a variety of objects and describing them in natural language. 
We injected MLLM into the system to periodically provide comprehensive information about the surroundings to inform blind users during exploration. 
In this paper, we investigate the appropriate presentation format, such as content types and lengths, and the quality of the responses from MLLMs through our user studies.

\subsection{Interaction} 
The ability for users to select destinations and routes according to their interests, often referred to as autonomy, is particularly important for exploration~\cite{Kaniwa2024ChitChatGuide,kayukawa2022HowUsers}. 
In our system, to what extent users prefer to take control over the robot (\ie, interaction) remains unknown.
Based on the scene descriptions given by the system~\cite{Kaniwa2024ChitChatGuide}, some blind users may fully embrace letting the robot guide them automatically, while others may prefer to decide which way to go on their own.
Additionally, this preference may also be influenced by the robot's descriptions of the scenes. 
Given that the extent of user preference for autonomy remains unclear, we first conducted the formative study (Sec.~\ref{sec:study1}) to explore the requirement of autonomy based on interaction needs. 
Then, we conducted a full study (Sec.~\ref{sec:study2}) to evaluate the users' opinions on autonomy in our improved system, which integrated the feedback from the formative study.

\vspace{-2mm}
\section{Formative Study}
\label{sec:study1}
We first conducted a formative study to investigate the requirements of the system, such as how the system should explain its surroundings and what potential interactions may happen between the robot and the user.
To conduct the study, we recruited ten participants through our existing email list.
Interestingly, our recruitment emails were shared among blind people, eventually reaching people not on our emailing list. 
In the recruitment email, we specified that those who are unfamiliar with the experimental location, \ie, even if they have had previous visiting experience, they do not have a clear understanding of the building or know what is there, would be eligible to participate.
Tab.~\ref{tab:demographics1} shows the demographics of the participants. 
All studies in this paper have been approved by our institution's review board.
Informed consent was read out to all participants in this paper and obtained from them. 
The study took approximately two hours, and the participants were compensated \$20 per hour and reimbursed for their transit costs.
Only one participant was present for each session. 

\begin{figure*}
    \centering
    \includegraphics[width=1\linewidth]{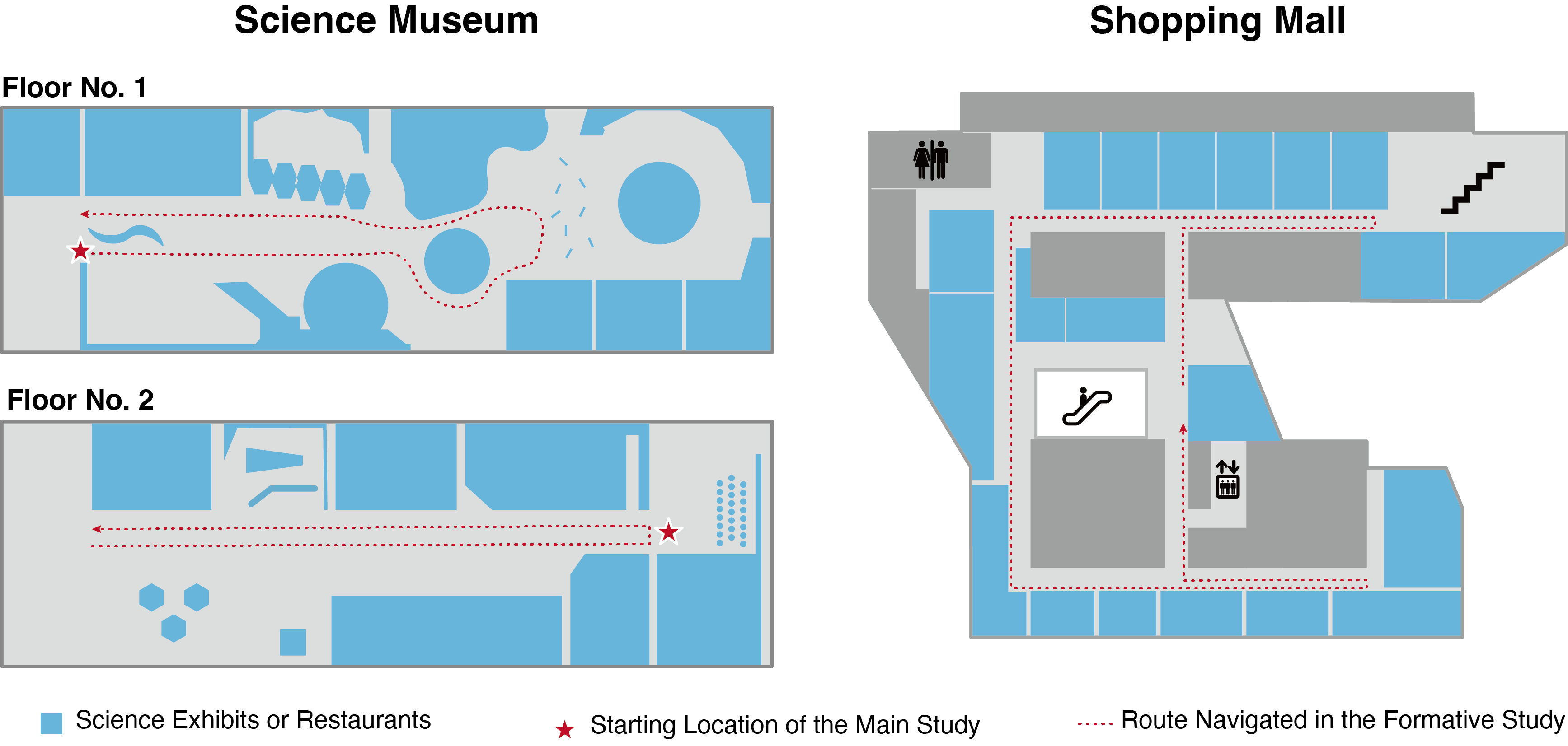}
    \caption{Floor maps of the location of the study. The left panel shows the two floors of the science museum, \rrred{the fifth floor of Miraikan}, which feature exhibits on various topics, such as environmental issues and space exploration. On the right panel is a floor plan of a shopping mall, \rrred{the fourth floor of Toranomon Hills Station Tower}, which includes a variety of restaurants offering different cuisines, including French, Japanese, Chinese, and cafes.}
    \Description{
    The image contains three floor plans, each marked with dotted red lines and stars, indicating different routes and destinations. On the left side are two floors of a science museum, while the right side displays a single-floor plan of a shopping mall. The first map of the science museum highlights exhibits in light blue, connected by hallways in light gray. The floor has a rectangular shape with a main pathway running through the center. There are three round exhibits—two in the middle and one on the right—with additional exhibits placed along the left and right sides of the main walkway. A red dotted line shows the route followed in the first study, starting from the left side, circling the floor, and returning to the starting point. The starting point for the main study is located at the same spot as in the formative study. The second map represents the museum's second floor, also rectangular in shape and similarly structured to the previous floor but without the round exhibits. Again, a red dotted line marks the path followed in the formative study, beginning on the left side, looping around the floor, and returning to the starting point. However, the main study on this floor begins at a different location, on the right side of the floor. The last map on the right shows part of a shopping mall, featuring shops in blue and common areas in gray. The floor has a maze-like layout with several intersections and includes 21 restaurants. The route for the formative study begins in the center of the floor, loops around each restaurant, and returns to the starting point.
    }
    \label{fig:route}
\end{figure*}

\subsection{Prototype System}
We developed our prototype robot system according to Sec.~\ref{sec:system_design}. 
It was based on an open-source robot platform\footnote{https://github.com/CMU-cabot/cabot} and could guide users while explaining the surrounding environment. 
To ensure that the participants experienced the same level of autonomy, we used teleoperation, a Wizrd-of-Oz-based approach~\cite{riek2012wizard-of-oz}, to force the robot to be in full-automatic mode when guiding the participants.
We adopted a suitcase-shaped wheeled robot for this study.
The suitcase's appearance allows blind users to seamlessly blend into their environment, leading to higher social acceptance from users, surrounding pedestrians, and facility managers~\cite{kayukawa2022HowUsers}. 
As shown in Fig.~\ref{fig:device&ui}--A-1, the robot has a handle embedded with five buttons, a touch sensor beneath the handle, a 360$^\circ$ \red{Velodyne VLP-16 LiDAR sensor~\cite{Velodyne}} sensor, three RGBD cameras with resolutions of 640×360, \red{one RealSense D455 camera~\cite{RealSenseD455} mounted at the front, two RealSense D435 cameras~\cite{RealSenseD435} on the left and right}, and a pair of motorized wheels in differential drive configuration.
\red{
Inside the suitcase, it has Ruby R8 powered by an AMD Ryzen R7-4800U CPU~\cite{NUC}, and a Jetson Mate featuring multiple Jetson Xavier NX GPUs~\cite{JetsonMate}.
}
The RGBD cameras were attached 0.51 meters above the ground.
The touch sensor detects whether or not the user is holding the handle and moves only when it is being held by the user. 
The cameras combined have a horizontal field of view of approximately 180$^\circ$.
The weight of the robot is approximately 15kg.
We set the default speed of the robot to 0.5 meters per second to maintain a balance between a comfortable walking speed and a speed that allows sufficient time to absorb the scene description audios.
A smartphone is attached to the suitcase to provide audio feedback through a neck speaker worn by users, connected via Bluetooth.

To convey the surrounding information to the participants, we used GPT-4o~\cite{GPT4o}, a popular MLLM model.
We inputted the images from the three RGBD cameras into the MLLM model and asked the model to generate descriptions of the surrounding environment.
The robot was designed to describe surrounding information 5-10 seconds after the end of the previous description every time. 
We engineered the prompts to ask the MLLM model to first provide a general overview of the scene, followed by specific details on the left, front, and right. 
We asked the descriptions to include as many objects as possible and incorporate layout information, such as navigable directions and the presence of walls~\cite{jain2023want}. 
\red{The processing time and cost to generate a description was 6.087 seconds and \$0.00740 on average.}
We attach the full prompts in Appendix Sec.~\ref{appendix:prompt_formative}.

\begin{table*}[]
\caption{Demographics of participants who attended the formative study. The table reports their gender, age, navigation aid, which they frequently use, frequency of exploration done either independently or with sighted people per year, their experimental location, number of previous visits to the experimental location, and analyzed preference. }
\Description{The table presents demographic data for participants in the formative study. The information includes gender, age, type of mobility aid used, the age at which they started using the aid, frequency of exploration per year, the location of the experiment, the number of previous visits, and their preference analysis. The following describes each participant. P01 is a 64-year-old female who uses a cane, started using it at the age of 44, explores 48 times per year, participated at the Science Museum with 1 previous visit, and is exploration-inclined. P02 is a 53-year-old male who uses a cane, started using it at the age of 13, explores 36 times per year, participated at the Science Museum with 0 previous visits, and is destination-oriented. P03 is a 74-year-old male who uses a cane, started using it at the age of 0, explores 1 time per year, participated at the Science Museum with 0 previous visits, and is destination-oriented. P04 is a 54-year-old female who uses a cane, started using it at the age of 0, explores 12 times per year, participated at the Science Museum with 0 previous visits, and is exploration-inclined. P05 is a 56-year-old male who uses a cane, started using it at the age of 52, explores 2 times per year, participated at the Science Museum with 0 previous visits, and is intermediate in preference. P06 is a 32-year-old male who uses a cane, started using it at the age of 0, explores 12 times per year, participated at a shopping mall with 0 previous visits, and is intermediate in preference. P07 is a 55-year-old female who uses a cane, started using it at the age of 52, explores 0 times per year, participated at a shopping mall with 1 previous visit, and is exploration-inclined. P08 is a 63-year-old male who uses a cane, started using it at the age of 22, explores 12 times per year, participated at a shopping mall with 0 previous visits, and is intermediate in preference. P09 is a 78-year-old female who uses a guide dog, started using it at the age of 22, explores 12 times per year, participated at a shopping mall with 0 previous visits, and is destination-oriented. P10 is a 49-year-old female who uses a cane, started using it at the age of 3, explores 1 time per year, participated at a shopping mall with 0 previous visits, and is exploration-inclined.}
\label{tab:demographics1}
\resizebox{\textwidth}{!}{%
\begin{tabular}{ccccccccc}
\toprule
    & Gender & Age & Aid       & \begin{tabular}[c]{@{}c@{}}Age \\  of Onset\end{tabular} & \begin{tabular}[c]{@{}c@{}}Frequency of\\  Exploration per Year\end{tabular} & \begin{tabular}[c]{@{}c@{}}Experiment \\  Location\end{tabular} & \begin{tabular}[c]{@{}c@{}}Number of \\  Previous Visits\end{tabular} & Preference Analysis     \\
    \midrule
P01 & F      & 64  & Cane      & 44                                                       & 48                                                                           & Science Museum                                                  & 1                                                                     & Exploration-Inclined \\
P02 & M      & 53  & Cane      & 13                                                       & 36                                                                           & Science Museum                                                  & 0                                                                     & Destination-Oriented \\
P03 & M      & 74  & Cane      & 0                                                        & 1                                                                            & Science Museum                                                  & 0                                                                     & Destination-Oriented \\
P04 & F      & 54  & Cane      & 0                                                        & 12                                                                           & Science Museum                                                  & 0                                                                     & Exploration-Inclined \\
P05 & M      & 56  & Cane      & 52                                                       & 2                                                                            & Science Museum                                                  & 0                                                                     & Intermediate         \\
P06 & M      & 32  & Cane      & 0                                                        & 12                                                                           & Shopping Mall                                                   & 0                                                                     & Intermediate         \\
P07 & F      & 55  & Cane      & 52                                                       & 0                                                                            & Shopping Mall                                                   & 1                                                                     & Exploration-Inclined \\
P08 & M      & 63  & Cane      & 22                                                       & 12                                                                           & Shopping Mall                                                   & 0                                                                     & Intermediate         \\
P09 & F      & 78  & Guide dog & 22                                                       & 12                                                                           & Shopping Mall                                                   & 0                                                                     & Destination-Oriented \\
P10 & F      & 49  & Cane      & 3                                                        & 1                                                                            & Shopping Mall                                                   & 0                                                                     & Exploration-Inclined \\
\bottomrule
\end{tabular}
}
\end{table*}

\subsection{Experimental Location}
To ensure the diversity of the findings we would obtain from this study, we conducted the study in two different locations. 
\red{We chose to conduct our studies in a science museum and a shopping mall, as these are locations where people typically engage in exploration, and they have been utilized in previous research~\cite{asakawa2019independent,asakawa2018present,Kaniwa2024ChitChatGuide}. 
A museum is generally a place for learning about exhibits, while a shopping mall often requires exploration both before and during visits to stores.
\rrred{Specifically, we used the fifth floor of Miraikan\footnote{\url{https://www.miraikan.jst.go.jp/en/}} for the science museum and the fourth floor of Toranomon Hills Station Tower\footnote{\url{https://www.toranomonhills.com/}} for the shopping mall.}
The floor map of the science museum is illustrated in the left panel of Fig.~\ref{fig:route}, which contains two floors, both primarily featuring science exhibits. 
For the studies, the order of the two floors was counterbalanced.
The study in the museum was conducted after business hours, during which customers were absent, but staff were present for their duties.
The floor map of the shopping mall is illustrated in the right panel of Fig.~\ref{fig:route}, a floor that contains several restaurants from various countries. 
The study in the shopping mall was conducted during regular business hours.}
As shown in Tab.~\ref{tab:demographics1}, the study with P01--P05 took place in the science museum, and the study with P06--P10 took place in the shopping mall.

\subsection{Procedure}
\begin{figure*}
    \centering
    \includegraphics[width=1\linewidth]{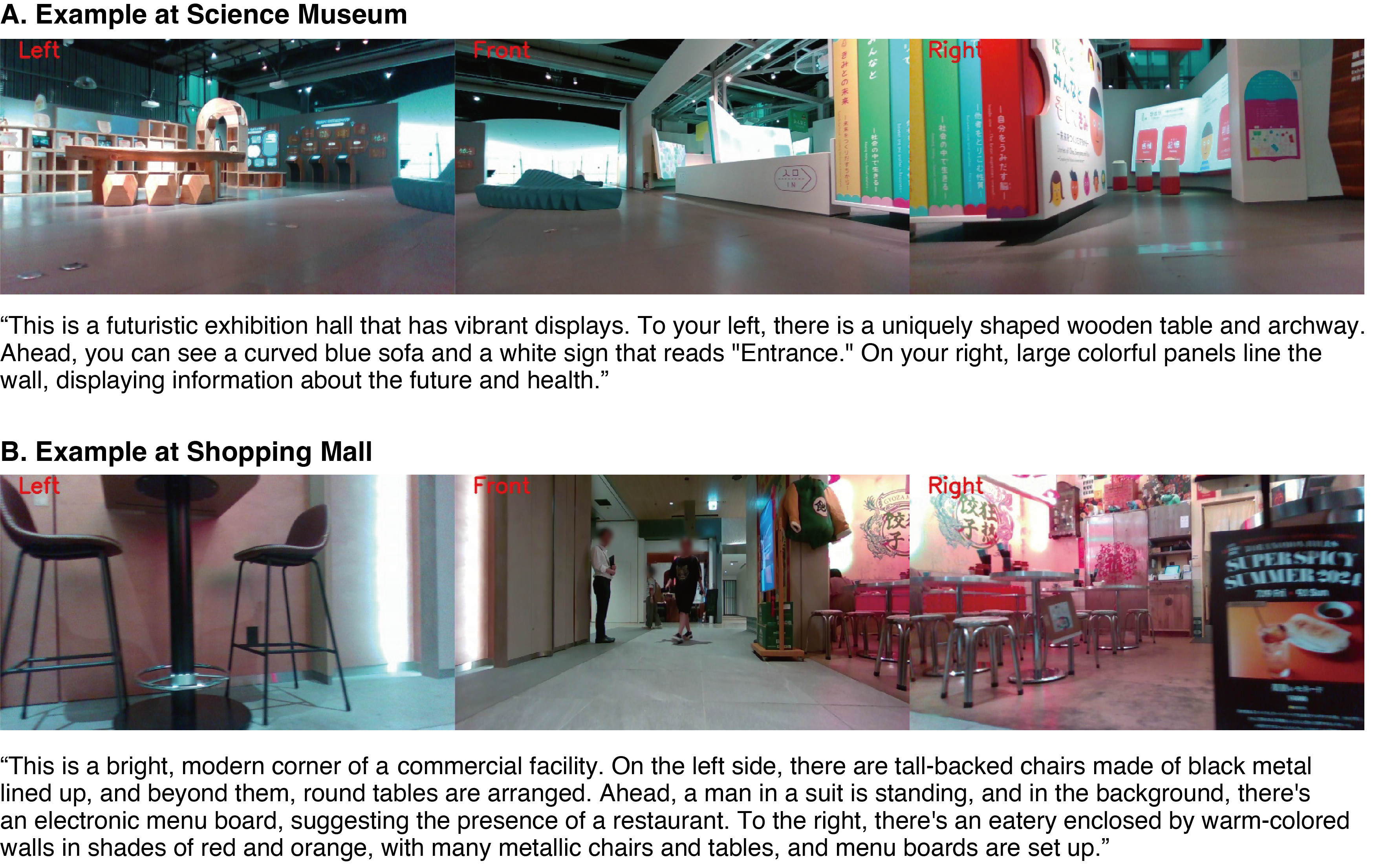}
    \caption{Examples of descriptions described in the formative study. Panel A shows an example of a description generated at the science museum, and Panel B shows the one generated at a shopping mall.}
    \Description{Examples of descriptions described in the formative study. Panel A shows an example of a description generated at the science museum, saying "This is a futuristic exhibition hall that has vibrant displays. To your left, there is a uniquely shaped wooden table and archway. Ahead, you can see a curved blue sofa and a white sign that reads "Entrance." On your right, large colorful panels line the wall, displaying information about the future and health." and Panel B shows the one generated at a shopping mall, saying "This is a bright, modern corner of a commercial facility. On the left side, there are tall-backed chairs made of black metal lined up, and beyond them, round tables are arranged. Ahead, a man in a suit is standing, and in the background, there's an electronic menu board, suggesting the presence of a restaurant. To the right, there's an eatery enclosed by warm-colored walls in shades of red and orange, with many metallic chairs and tables, and menu boards are set up."}
    \label{fig:study1example}
\end{figure*}

For each participant, we first conducted a pre-study interview to learn about their experience in exploring buildings, followed by an explanation that the study aimed to gather their opinions on a guide system designed to assist with exploration.
Then, participants were given a task to navigate the predetermined route (red arrow of Fig.~\ref{fig:route}) guided by the robot.
Adopting a Wizrd-of-Oz-based approach, an experimenter controlled the robot to navigate along the route and stop when there were nearby pedestrians. 
During exploration, the robot periodically generated descriptions of the scenes. 
We show an example of the generated description in Fig.~\ref{fig:study1example}.
After the exploration, we asked the participants if there were any additional things they wanted to do to partially simulate the potential interaction, such as going to additional places or going around the floor again for more exploration.
Finally, we conducted a post-interview session to gather their feedback on the system. 

\subsection{Result}
\subsubsection{Interests to Exploration}
All participants stated that totally independent exploration is challenging, but they expressed a desire for exploration if a guide system can help them do so. For example:
\newanswer[\label{P02Conditioned}]\textit{``I don't really explore much. I go out with a specific purpose in mind [...] The reason is that it's just too bothersome. But I do think it would be fun if I did [...]  I'm more of an old-timer, so exploration never really caught my interest. It's not that I didn't care at all, but perhaps I've been living this way (not to explore).}\footnote{The comments were obtained in the native language where the study was conducted. We translate the comments into English using publicly available LLM to ensure reproducibility. We show the full prompt used for translation in Appendix Sec.~\ref{appendix:translate}.} (P02)

\subsubsection{Positive Feedback and Appreciated Information}
Seven participants (P01, P04--P08, and P10) expressed their enjoyment while navigating with the robot, particularly with the provided surrounding descriptions, as described in the following comment:
\newanswer[\label{P07Enjoy}]\textit{``My first impression was that it was a lot of fun. The reason is, as you just mentioned, unlike the person I usually walk with, the system provided detailed explanations about things like the color of the walls and the signs we saw and even described how the chef was preparing the food. Normally, you might get some of this information from others, but it's rare to get such thorough details. I found myself thinking, ``Oh, I see, that's how it looks to sighted people,'' and I felt there was a lot of new information. In that sense, I really enjoyed it.''} (P07)

Participants appreciated a variety of real-time details about their surroundings, notable examples include patterns on the walls, lighting conditions, subjective descriptors such as ``beautiful,'' the presence and actions of nearby people, the existence of signboards, the layout of the environment, and the visibility of a chef in an open kitchen. 
Additionally, P10, who requested to walk around the floor again, noted that receiving different descriptions of the same location was beneficial, as it gave them a sense of presence:
\newanswer[\label{P10VariousAndDifferentInformation}]\textit{``The system mentioned those things, as well as details about the plants and wall decorations. It's like, you talked about so many different things that it feels like I was actually looking around myself. Honestly, most of the time, I get so occupied with just reaching my destination that I don't notice things around me. [...] The system also mentioned things in the second round of explanations that weren't covered in the first round, which was nice. It conveyed a sense of the ongoing atmosphere and gave a good understanding of the situation at the time.''} (P10)

\subsubsection{Information Needs}
\label{sec:info_needs}
Participants hoped for further polishing of the delivered information about the scenes. Six participants (P01--P03, P06, and P09--P10) felt the information conveyed about the surroundings was too abstract, indicating the need for more specific information:
\newanswer[\label{P01NeedMoreConcreteInformation}]\textit{``The system talked about there are just exhibits, or there's information on panels, but I think it would be nice if the system talked about specific titles. There are places where the system talked about them, but there are also places where it did not, so I found myself wondering about that.''} (P01)
In particular, three participants (P02, P03, and P09) commented that the descriptions neither helped them learn the environment nor make decisions such as determining which shops or exhibits to enjoy:
\newanswer[\label{P09NegativeImpression}]\textit{``I expected it to at least tell me the name of the store, but it was disappointing to find out that it didn't do that at all. I really wish there was a system that could provide pinpointed information about what I want to know. Especially in an unfamiliar restaurant area, for example, if I come alone and use the device to enter the premises, it starts running, and then when I think, ``Oh, should I have Japanese food today, or maybe tonkatsu?'', without such information, I end up just walking around aimlessly.''} (P09)

Participants also described specifics about what types of information would be beneficial to include, such as the position of objects given in meters and clock directions, the availability of seats, people on collision paths, identities of surrounding individuals (\eg staff), and specific names of objects. 
In science museums, participants also wanted to know whether exhibits are touchable. 
In shopping malls, participants also wanted to learn the store menus and whether there is a spacious area for a guide dog to rest while the user is eating.
However, three other participants (P02, P03, and P09) found certain information, such as details about lighting, surrounding people, and wall design, unnecessary. 

\subsection{Design Considerations}
The results of the study affirmed that there are certain appreciations and room for improvement for the exploration robot for blind people.
Based on the above results, we derived several requirements for the system, as listed below.

\subsubsection{Vary Detail of Descriptions Based on Preferences and Contexts}
\label{sec:implication_varydetail}
We observed three types of preferences: one that enjoyed all the descriptions provided by the system (\textit{Exploration-Inclined}), another that enjoyed the descriptions but preferred to limit certain information (\textit{Intermediate}), and a third group that only wanted information useful for determining where to go (\textit{Destination-Oriented}). 
In Tab.~\ref{tab:demographics1}, we show the description preference of each participant.
To classify the preferences, we first classified three participants who did not enjoy the description of the system as \textit{Destination-Oriented}.
Then, based on the discussion between the authors, we classified the rest as \textit{Intermediate} or \textit{Exploration-Inclined}.
Furthermore, the type of information needed varied slightly depending on the experimental location. 
For instance, participants sought seating information for guide dogs in shopping areas, whereas in the science museum, they were more interested in whether the exhibits were touchable.
Given these three types of preferences and context-dependent information needs, we modified the system so that it could adjust the amount and types of information conveyed to each participant.

\subsubsection{Add Question and Answer Functionality}
\label{sec:implication_Q&A}
There was a clear need for question-and-answer (Q\&A) interaction, as seven participants (P02--P05 and P08--P10) noted that they would like the option to ask more detailed questions through conversation. 
Participants expressed interest in this functionality when they were curious about the system's descriptions. This would allow them to ask more detailed questions about the objects of interest.

\subsubsection{Add ``Take-Me-There'' Functionality} 
\label{sec:implication_takemethere}
Four participants (P02, P04, P06, and P10) mentioned that they would like to revisit locations they found interesting after walking around the floor. 
Example situations include deciding to visit a shop, engaging with touchable exhibits, or returning to chairs discovered during the exploration. 
In unfamiliar locations, where users may lose their sense of direction, participants also expressed the need for a feature that guides them back to their initial location~\cite{kuribayashi2023pathfinder}.

\subsubsection{Vary Speed and Be Able to Stop the Robot}
\label{sec:implication_speed}
While the majority found the default speed appropriate for listening and understanding the described information, there were requests for customizable speed settings. 
Eight participants stated that the robot's speed was appropriate for exploring. 
Two participants (P04 and P06) expressed a preference for a faster speed.
P01 additionally wanted to stop when the robot read out the descriptions of interest.
In conclusion, users who are \textit{Destination-Oriented} or have already determined the destination through exploration may want to increase the speed, while users who prefer to take time exploring might wish to slow down or stop the robot entirely. 

\subsubsection{Add Direction Specifying Functionality}
\label{sec:implication_directionspecification}
Participants expressed a desire for more active engagement by specifying the movement direction themselves. 
Four participants (P02--P05) mentioned that they wanted more active control over the movement direction based on their interests.
Additionally, we extrapolated that instead of simply following the robot, some users may prefer to interactively choose the direction based on the audio description of the surroundings.
This could lead to greater autonomy because it would enrich the exploratory experience by aligning the robot's movement with the users' real-time curiosity and needs, creating a more personalized and engaging exploration experience.

\section{WanderGuide Implementation}

\label{sec:Implementation}
\red{
In this section, we provide the implementation of WanderGuide informed by the formative study.
Below is a summary of updates made from the implementation of the formative study.
\begin{itemize}
\item Attachment of a new fisheye camera for a better view (Sec.~\ref{sec:device_update})
\item Implementation of a waypoint detection algorithm for realizing autonomous map-less navigation (Sec.~\ref{sec:waypoint_detection})
\item Implementation of three levels of description based on user preferences (Sec.~\ref{sec:Implementation_description_mode})
\item Implementation of ``Take-Me-There'' Functionality (Sec.~\ref{sec:take_me_there})
\item Implementation of two navigation modes automatic navigation mode and manual control mode (Sec.~\ref{sec:implementation_button})
\item Implementation of an interface to adjust speed, level of description, and navigation mode (Sec.~\ref{sec:implementation_button})
\item Implementation of Q\&A Functionality (Sec.~\ref{sec:QA})
\end{itemize}
}

\subsection{Hardware Update} 
\label{sec:device_update}
One of the notable user feedbacks was the need for more detailed information, such as the names of POIs.
However, the cameras in the prototype system were mounted at only 0.51 meters above the ground, had low resolution, and had a limited vertical field of view, making it difficult for the MLLM model to consistently capture details.
Thus, as illustrated in Fig.~\ref{fig:device&ui}--A-2, we attached a fisheye camera with 1080p resolution and a wide field of view to the higher part of the robot.

\subsection{Waypoint Detection and Navigation}
\label{sec:waypoint_detection}

\begin{figure*}
    \centering
    \includegraphics[width=\linewidth]{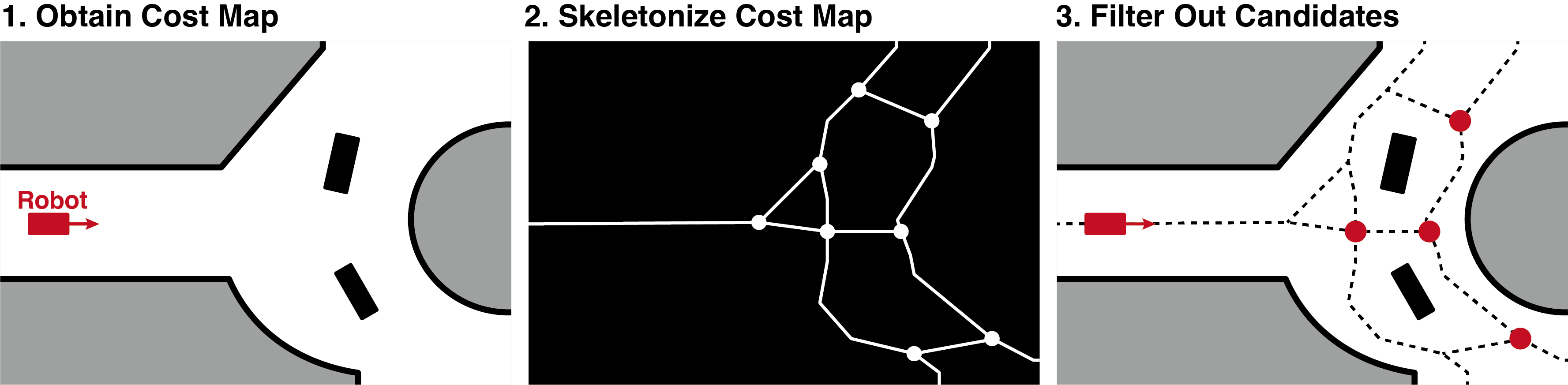}
    \caption{Three steps of the waypoint detection algorithm. Step 1 shows the generated cost map, while Step 2 depicts the skeletonization process of the cost map along with the detection of intersection points. Finally, Step 3 highlights the selected intersection points, which are identified as waypoint candidates.}
    \Description{
    The image depicts three stages of a waypoint detection algorithm. On the left, Step 1 shows a simplified map where a red rectangular robot is moving toward an open area. The map highlights obstacles in black and the surrounding areas in gray. The white space represents the robot's navigable area. In the center, Step 2 illustrates a skeletonized version of the map from Step 1. Thin white lines trace the navigable paths, forming a network-like structure. At several points where the paths intersect, small white dots indicate intersection points. On the right, Step 3 marks the final waypoint candidates. The robot, represented by a red rectangle, is shown along with red dots that represent the selected waypoint candidates. Dotted black lines trace possible paths from the robot toward these waypoints, navigating around the obstacles.
    }
\label{fig:waypoint_detection}
\end{figure*}

In order to produce destinations to navigate to for the users, a waypoint detection algorithm (Fig.~\ref{fig:waypoint_detection}) is necessary to determine navigable points for the robots.
As no prebuilt maps were available, we first constructed a cost map, a two-dimensional occupancy grid that assigns costs based on obstacles, and updated it in real-time. 
We utilized the \red{existing open-source Cartographer package~\cite{Cartographer}, which is a real-time Simultaneous Localization and Mapping (SLAM) algorithm,} to generate the cost map.
Next, the cost map was skeletonized, and intersection points on the skeleton were identified based on a kernel-based corner detection algorithm~\cite{soille1999morphological}.
The intersection points, which are typically far from obstacles, were next used to select potential waypoints.
To maintain sparsity among waypoints, we applied the DBSCAN clustering algorithm~\cite{ester1996density} over the intersection points, and selected the centers of the clusters as potential waypoints.
In addition, coordinates three meters in front, behind, and to the sides of the robot were also considered potential destinations to address the case where no intersection points were detected through the algorithm.
As selecting a waypoint too far may be challenging for the robot to find a suitable path and a waypoint too close would lead to frequent destination changes, we filtered out candidates further than 50 meters and closer than one meter to the robot.
After filtering, the final list of candidate waypoints was set.

During navigation, the robot automatically selected its goal from the candidate waypoints. 
By default, priority was given to waypoints lying in the same forward direction as the robot's initial orientation, where it was placed and activated. 
If no forward waypoints are available, the robot selects the waypoint with the smallest absolute angle relative to its current orientation. 
Once a waypoint was chosen from the candidate list, the robot navigated to it by using the onboard open-source navigation algorithm~\cite{guerreiro2019cabot}.
Once the waypoint was reached, the next waypoint was chosen automatically using the same process.
It is important to note that prioritizing the robot's initial orientation was based on the assumption that users can adjust the general direction to proceed, such as starting from the entrance into the building.

\subsection{Scene Description Generation}
\label{sec:Implementation_description_mode}
The basic algorithm for scene description generation remains unchanged, but the description was conveyed only when the robot was moving.
Also, the MLLM took the overall view image from the fisheye view camera in addition to the three RGB images from the RGBD cameras.
According to the results of the formative study, we added three levels of detail in the scene description.
\begin{itemize}
    \item{\textit{Detailed Description}}
    This mode provided rich, immersive descriptions for blind users who wanted to explore their surroundings in detail. 
    The MLLM generated 3-4 sentences (120-240 characters), covering lighting, signs, layout, nearby people, and subjective descriptors like ``beautiful'' or ``modern''. 
    The description began with an overview, followed by details of the left, front, and right.
    \item{\textit{Balanced-Length Description}}
    This mode offered clear descriptions for users who preferred concise but informative content. 
    The MLLM generated 2-3 sentences (60-120 characters), focusing on relevant details like signs and layout, while omitting lighting conditions or subjective descriptors. 
    Descriptions covered the left, front, and right, without the overview.
    \item{\textit{Concise Description}}
    This mode provided brief, essential information for users who wanted quick guidance. 
    The MLLM generated 1-2 sentences (less than 60 characters), focusing only on key details needed to navigate, excluding unnecessary information. 
    Descriptions covered the left, front, and right, without the overview.
\end{itemize}
For MLLM, these three levels were controlled via prompts, which are shown in Appendix Sec.~\ref{appendix:prompt_main}.
All prompts shared the following instructions in common: to convey environmental information that assists blind people to explore, to refer to specific details such as genres or the names of objects, to encourage reading any text that helps users explore, to describe spaces for guide dogs to sit in restaurants, to provide information about potential hazards, and to use numbers to indicate the relative positions of surrounding objects.
To ensure that the MLLM adhered to the instructions provided in the prompt, we employed a two-stage inference process. 
First, we instructed MLLM to perform an initial inference, generating a description. 
Then, it self-supervised this generated description to verify if it met the given instructions. 
Finally, MLLM produced a revised version of the description to be presented to the user.
Although this approach resulted in longer inference times, the outputs produced follow complex prompt instructions.
\red{
The description is read aloud every 5-10 seconds after the previous description has been read out.
The processing time and cost to generate a description was 5.78 seconds and \$0.00811 for a Detailed Description, 4.75 seconds and \$0.00753 for a Balanced-Length Description, and 4.02 seconds and \$0.00734 for a Concise Description on average.
}

\subsection{``Take-Me-There'' Functionality}
\label{sec:take_me_there}
Acting on the feedback received from the formative study, we implemented a function that guided users to a destination verbally specified.
This feature was typically enabled by the robot's \textit{semantic map}~\cite{shafiullah2022clip,yokoyama2024vlfm,liu2024dragon}. 
In our case, we linked the images and generated descriptions, which had been saved as the robot had navigated, to the cost map of the robot.
Given a verbal cue from the user (\eg \textit{``I want to go to the blue sofa.''}), the system first used our selected MLLM model to extract the name of the target location (\eg, blue sofa).
Then, we calculated the embeddings of the target location, all saved captured images, and all saved generated descriptions. 
We took a dot-product similarity between the extracted target location and the embeddings of images and descriptions to find the closest match.
We used pre-trained feature extraction models: a fine-tuned SimCSE~\cite{gao2021simcse} model for generating sentence embeddings from text and a pre-trained CLIP~\cite{radford2021learning} model for creating image embeddings.
We used models that were trained in the native language where the study was conducted.
The coordinate linked to the closest matched image or description would be set as the destination.
If the user wanted to go back to the initial location, we used MLLM to detect the user's intent and set the destination to the initial point.
\red{
We note that a similar functionality, the ``Take-Me-Back'' functionality, which allows users to return to their initial location, has been implemented in the previous map-less navigation system PathFinder~\cite{kuribayashi2023pathfinder}. The ``Take-Me-Back'' functionality is specifically designed for navigation purposes, as it was motivated by the challenge blind individuals face in returning to their original location after navigating. In contrast, our functionality is tailored for exploration tasks, enabling users to return to any point of interest they identified during their exploration. Ultimately, our functionality encompasses the capabilities of the ``Take-Me-Back'' feature while extending its application to support exploratory activities.
}

\begin{table*}[]
\caption{Demographics of participants who attended the main study. The table reports their gender, age, navigation aid, which they frequently use, frequency of exploration done either independently or with sighted people per year, their experimental location, and number of previous visits to the experimental location. }
\Description{
The table presents demographic information for participants in the main study, including gender, age, mobility aid used, the age at which they started using the aid, frequency of exploration per year, experiment location, the number of previous visits, and usability and workload metrics (SUS and Raw-TLX). The following describes each participant. P11 is a 59-year-old male who uses a cane, started using it at the age of 29, does not explore per year, participated at the Science Museum with 1 previous visit. P12 is a 59-year-old female who uses a cane, started using it at the age of 43, explores 36 times per year, participated at the Science Museum with 0 previous visits. P13 is a 56-year-old male who uses a cane, started using it at the age of 45, explores 1 time per year, participated at the Science Museum with 0 previous visits. P14 is a 60-year-old female who uses a cane, started using it at the age of 45, explores 48 times per year, participated at the Science Museum with 1 previous visit. P15 is a 59-year-old male who uses a cane, started using it at the age of 24, explores 4 times per year, participated at the Science Museum with 0 previous visits.
}
\label{tab:demographics2}
\begin{tabular}{cccccccc}
\toprule
    & Gender & Age & Aid  & \begin{tabular}[c]{@{}c@{}}Age \\  of Onset\end{tabular} & \begin{tabular}[c]{@{}c@{}}Frequency of\\  Exploration per Year\end{tabular} & \begin{tabular}[c]{@{}c@{}}Experiment \\  Location\end{tabular} & \begin{tabular}[c]{@{}c@{}}Number of \\  Previous Visits\end{tabular}\\
    \midrule
P11 & M      & 59  & Cane & 29                                                       & 0                                                                            & Science Museum                                                  & 1                \\
P12 & F      & 59  & Cane & 43                                                       & 36                                                                           & Science Museum                                                  & \red{1}                \\
P13 & M      & 56  & Cane & 45                                                       & 1                                                                            & Science Museum                                                  & 0                \\
P14 & F      & 60  & Cane & 45                                                       & 48                                                                           & Science Museum                                                  & 1                \\
P15 & M      & 59  & Cane & 24                                                       & 4                                                                            & Science Museum                                                  & 0                \\
\bottomrule
\end{tabular}
\end{table*}

\subsection{Navigation Mode and User Interface}
On the high level, we implemented button controls and conversation interaction methods for users to interact with the robot.

\subsubsection{Button Controls}
\label{sec:implementation_button}
\red{
We utilized the four directional buttons and the central button on the handle of the suitcase-shaped robot to enable users to control the robot's speed, adjust the level of descriptions, switch between automatic and manual control modes, and specify the direction of movement.
}
The mapping of the buttons is illustrated in Fig.~\ref{fig:device&ui}--B-1 and B-2.
The central button was used for mode changes.
The functions of the directional buttons would change depending on the robot's modes:  \textit{auto mode}, \textit{manual control mode}, and \textit{conversation mode}.
In auto mode, the robot navigated by determining the waypoint automatically.
The left and right buttons allowed the user to switch between three levels of description, where the default mode is the balanced-length description mode. 
The forward and backward buttons were used to adjust the robot’s speed.
Users can adjust the speed from zero to one meter per second, with increments of 0.05 meters per second.
In manual mode, users could specify directions on their own.
The robot would instruct the user to press the directional buttons to select the direction to proceed.
If there was a suitable waypoint in the specified direction, the robot would inform the users via voice feedback.
Otherwise, the robot conveyed that there were no navigable points in the specified direction.
In conversation mode, triggered by long-pressing the central button, the robot would pause, and all four directional buttons were disabled until the conversation was ended.
Users could manually end the conversation by long-pressing the central button again.
The details of the conversation mode are described below.

\subsubsection{Conversation}
\label{sec:QA}
The conversation mode allowed users to give commands or ask questions with verbal input via the smartphone attached to the robot (Fig.~\ref{fig:device&ui}).
When the user inputted their verbal cue, the system used MLLM to classify the user's intent into one of three categories: usage of ``Take-Me-There'' functionality, usage of Q\&A functionality, and direction specification.
If the detected intent was direction specification (\eg, \textit{``I want to go to right''}) the robot would navigate to the waypoint in the specified direction accordingly.
Finally, users could finish the conversation with an ending phrase such as \textit{``Thank you.'' }

\section{Main User Study}
\label{sec:study2}
This study was conducted to validate WanderGuide and explore further design space.
Participants were recruited and compensated similarly to those in the formative study.
Similar to the formative study, in the recruitment email, we specified that participants unfamiliar with the experimental location would be eligible to participate.
We conducted this study on the same two floors of the science museum. 
Tab.~\ref{tab:demographics2} shows the demographics of the participants.
\red{
None of the participants from the formative study participated in this study.
Similar to the formative study, this study was conducted after business hours.
}

\subsection{Task and Procedure}
For each participant, we first conducted a pre-study interview similar to the formative study.
Then, the participant joined a 30-minute training session to get familiar with the robot system before the main tasks.
For the main tasks, they were asked to freely explore the floor for 20 minutes using the system from a fixed starting location, as illustrated in Fig.~\ref{fig:route}.
The ordering of the floors was counterbalanced to mitigate the order effect.
After the main tasks, we conducted a post-study interview to ask several seven-point Likert scale questions (1: Strongly Disagree, 4: Neutral, and 7: Strongly Agree) that measure their self-evaluated exploration performance, Raw Task Load Index (TLX)~\cite{byers1989traditional} to measure the task workload, and system usability scale (SUS)~\cite{brooke1996sus} to evaluate the usability of the system.
Finally, we asked open-ended questions to gather comments on the system.
Below, we report the results of the study.

\begin{table}[]
\caption{The statistics of duration time and the count of interactions for each mode (Auto, Conversation, and Manual Control).
The ratio of the duration time is calculated based on the total duration time of the experiment per participant.}
\Description{
The table provides an analysis of the duration ratio (percentage of total time) and the count of interactions for three different modes: Auto, Conversation, and Manual Control. The statistics are presented for five participants: P11, P12, P13, P14, and P15. P11 spent 59.77\% of the time in Auto mode with 25 interactions, spent 37.52\% of the time in Conversation mode with 21 interactions, and spent 2.70\% of the time in Manual Control mode with 4 interactions. P12 spent 91.66\% of the time in Auto mode with 13 interactions, spent 8.16\% of the time in Conversation mode with 9 interactions, and spent 0.18\% of the time in Manual Control mode with 1 interaction. P13 spent 67.88\% of the time in Auto mode with 12 interactions, spent 30.56\% of the time in Conversation mode with 9 interactions, and spent 1.56\% of the time in Manual Control mode with 1 interaction. P14 spent 64.86\% of the time in Auto mode with 17 interactions, spent 33.94\% of the time in Conversation mode with 15 interactions, and spent 1.20\% of the time in Manual Control mode with 1 interaction. P15 spent 58.53\% of the time in Auto mode with 28 interactions, spent 20.03\% of the time in Conversation mode with 19 interactions, and spent 21.44\% of the time in Manual Control mode with 10 interactions.
}
\label{tab:activity_breakdown}
\begin{tabular}{@{}lcccccc@{}}
\toprule
    & \multicolumn{2}{c}{Auto} & \multicolumn{2}{c}{Conversation} & \multicolumn{2}{c}{Manual Control} \\
    & Ratio(\%)          & Count         & Ratio(\%)          & Count         & Ratio(\%)           & Count          \\ \midrule
P11 & 59.77         & 25            & 37.52         & 21            & 2.70           & 4              \\
P12 & 91.66         & 13            & 8.16          & 9             & 0.18           & 1              \\
P13 & 67.88         & 12            & 30.56         & 9             & 1.56           & 1              \\
P14 & 64.86         & 17            & 33.94         & 15            & 1.20           & 1              \\
P15 & 58.53         & 28            & 20.03         & 19            & 21.44          & 10             \\ \bottomrule
\end{tabular}
\end{table}
\begin{table}[]
\caption{Analysis of participants' requests to the system during conversation mode. We defined three types of queries, General Query, Specific Query, and Command Query, and classified each participant request into one of them. The ratios here are simply the count percentages over total counts.}
\Description{
The table provides a analysis of participants' requests made to the system during conversation mode. Requests are categorized into three types: General Query, Specific Query, and Command Query. The ratio (percentage) and count of each type of request are shown for each participant, along with the total number of requests. The following summarizes the data for each participant: P11 made 11.43\% of their requests as General Queries (4 requests), 45.71\% as Specific Queries (16 requests), and 42.86\% as Command Queries (15 requests), with a total of 35 requests. P12 made 30.00\% of their requests as General Queries (3 requests), 10.00\% as Specific Queries (1 request), and 60.00\% as Command Queries (6 requests), with a total of 10 requests. P13 made 61.54\% of their requests as General Queries (8 requests), 38.46\% as Specific Queries (5 requests), and 0.00\% as Command Queries (0 requests), with a total of 13 requests. P14 made 14.29\% of their requests as General Queries (4 requests), 46.43\% as Specific Queries (13 requests), and 39.29\% as Command Queries (11 requests), with a total of 28 requests. P15 made 8.33\% of their requests as General Queries (2 requests), 25.00\% as Specific Queries (6 requests), and 66.67\% as Command Queries (16 requests), with a total of 24 requests.
}
\label{tab:request_breakdown}
\resizebox{\columnwidth}{!}{%
\begin{tabular}{@{}cccccccc@{}}
\toprule
                     & \multicolumn{2}{c}{General Query} & \multicolumn{2}{c}{Specific Query} & \multicolumn{2}{c}{Command Query} & \multirow{2}{*}{Total}\\
                     & Ratio (\%)       & Count      & Ratio (\%)       & Count      & Ratio (\%)        & Count       \\ \midrule
P11  & 11.43            & 4         & 45.71            & 16         & 42.86              & 15           &  35\\ 
P12  & 30.00            & 3         & 10.00            & 1          & 60.00              & 6           &  10\\ 
P13  & 61.54            & 8         & 38.46            & 5          & 0.00              & 0           &  13\\ 
P14  & 14.29            & 4         & 46.43            & 13         & 39.29              & 11           &  28\\ 
P15  & 8.33            & 2         & 25.00            & 6         & 66.67              & 16           &  24\\ \bottomrule
\end{tabular}
}
\end{table}

\subsection{Analysis of Participants Activity During The Task}
\label{sec:activity_breakdown}
We report the statistics of each participant's activity during the task by referring to the system's log and the video captured during the tasks. 
Tab.~\ref{tab:activity_breakdown} shows the analysis of their time spent on the three modes as specified in Sec.~\ref{sec:implementation_button}.
We noticed that the activation quantity and duration of each mode varied significantly among participants.
P11, P13, P14, and P15 frequently used the conversation mode.
Notably, P11 spent nearly 40\% of the total time engaging in conversation with the robot.
In contrast, P12 barely used the conversation mode and relied on the auto mode for 90\% of the total time.
P15 was the only participant who actively used the manual control mode.

\begin{table*}[]
\caption{Error analysis of outputs from MLLM.  We classified the errors into five categories and counted the number of them. Note that a single response could contain multiple errors, so the sum of errors does not match the total output of MLLM.}
\Description{
The table presents an error analysis of outputs from an MLLM (Multimodal Large Language Model). The errors are categorized into five types: Wrong Character Recognition, Wrong Object Recognition, Nonexistent Objects and Texts, Misunderstanding User Input, and Inaccurate User Input. It also includes a count of outputs with no errors. The total number of outputs is also provided, and a single response can contain multiple errors. The following describes the findings. For Scene Description, there were 31 instances of Wrong Character Recognition, 6 instances of Wrong Object Recognition, 11 instances of Nonexistent Objects and Texts, no occurrences of Misunderstanding or Inaccurate User Input, 117 outputs with No Error, and a total of 164 outputs. For Q\&A Response, there were 9 instances of Wrong Character Recognition, 6 instances of Wrong Object Recognition, 15 instances of Nonexistent Objects and Texts, 5 instances of Misunderstanding User Input, 1 instance of Inaccurate User Input, 21 outputs with No Error, and a total of 53 outputs.
}
\label{tab:hallucinations}

\begin{tabular}{@{}cccccccc@{}}
\toprule
                  & \begin{tabular}[c]{@{}c@{}}Wrong\\ Character \\ Recognition\end{tabular} & \begin{tabular}[c]{@{}c@{}}Wrong\\ Object\\ Recognition\end{tabular} & \begin{tabular}[c]{@{}c@{}}Nonexistent\\ Objects and \\Texts\end{tabular} & \begin{tabular}[c]{@{}c@{}}Misunderstanding\\ User\\ Input\end{tabular} & \begin{tabular}[c]{@{}c@{}}Inaccurate\\ User\\ Input\end{tabular} & \begin{tabular}[c]{@{}c@{}}No\\ Error\end{tabular} & \begin{tabular}[c]{@{}c@{}}Total \\ output\end{tabular} \\ \midrule
Scene Description & 31                                                  & 6                                                                    & 11                                                                   & -                                                                     & -                                                          & 117                                                   & 164                                                          \\
Q\&A Response     & 9                                                   & 6                                                                    & 15                                                                   & 5                                                                     & 1                                                          & 21                                                    & 53                                                           \\ \bottomrule
\end{tabular}%
\end{table*}
\begin{table}[]
\caption{The statistics of the usage of each description level.
Usage statistics for each description level are calculated by normalizing the duration of each level with respect to the total duration of the experiment.
}
\Description{
The table presents the usage statistics for different description levels—Concise, Balanced-Length, and Descriptive—by normalizing the duration of each level with respect to the total duration of the experiment for five participants (P11 to P15). P11 used Concise descriptions 0.10\% of the time, Balanced-Length descriptions 76.46\% of the time, and Descriptive descriptions 23.43\% of the time. P12 used Concise descriptions 15.22\% of the time, Balanced-Length descriptions 29.88\% of the time, and Descriptive descriptions 54.91\% of the time. P13 used Concise descriptions 0.19\% of the time, Balanced-Length descriptions 49.14\% of the time, and Descriptive descriptions 50.66\% of the time. P14 used Concise descriptions 0.15\% of the time, Balanced-Length descriptions 87.94\% of the time, and Descriptive descriptions 11.91\% of the time. P15 used Concise descriptions 0.14\% of the time, Balanced-Length descriptions 99.86\% of the time, and Descriptive descriptions 0.00\% of the time.
}
\label{tab:description_level}
\begin{tabular}{@{}lccc@{}}
\toprule
    & Concise & Balanced-Length & Detailed \\ \midrule
P11 & 0.10\%  & 76.46\%         & 23.43\%     \\
P12 & 15.22\% & 29.88\%         & 54.91\%     \\
P13 & 0.19\%  & 49.14\%         & 50.66\%     \\
P14 & 0.15\%  & 87.94\%         & 11.91\%     \\
P15 & 0.14\%  & 99.86\%         & 0.00\%      \\ \bottomrule
\end{tabular}
\end{table}

\subsection{Analysis of Requests from Participants During Within The Conversation Mode}
\label{sec:request_breakdown}
In Tab.~\ref{tab:request_breakdown}, we further report the statistics of requests from participants within the conversation mode. 
Note that the total count of conversations in Tab.~\ref{tab:request_breakdown} is bigger than the conversation mode counts in Tab.~\ref{tab:activity_breakdown}, as multiple turns of conversation could happen in one conversation mode interaction.
We classify each verbal request into three categories. 
\begin{description}
    \item[General Query] Request general information in the surrounding area or in a particular direction.
    \item[Specific Query] Request detailed information about a specific object in the environment.
    \item[Command Query] Issue command to guide to destination, triggering ``Take-Me-There'' functionality or direction specification via conversation.
\end{description}

Overall, we discovered that although our system constantly provided environmental descriptions in auto mode, users still preferred to ask for general information about their surroundings or in a specific direction in conversation mode. 
For example, P13 predominantly made General Queries (61.54\%). 
Users also had diverse preferences when using our system. 
Some users such as P11 (45.71\%), P13 (38.46\%) and P14 (46.43\%) were interested in learning the specifics of POIs, reflecting the takeaways obtained in Sec.~\ref{sec:info_needs}. 
Some users such as P11 (42.86\%), P12 (60.00\%), P14 (39.29\%), and P15 (66.67\%) favored using conversation mode to instruct the robot to guide them to their destinations. 
In particular, by referencing Tab.~\ref{tab:activity_breakdown}, we can see that P11, P12, and P14 preferred conversation mode over manual control mode to issue commands. 
This validates the extrapolated idea in Sec.~\ref{sec:implication_directionspecification}.

\subsection{Error Analysis of Scene Description and Q\&A Responses}
\label{sec:error_analysis}
In Tab.~\ref{tab:hallucinations}, we report the accuracy of MLLM responses both during auto and conversation modes.
We manually analyzed the text output generated by MLLM and compared it with the logs of the images saved.
We classified and counted the errors made by MLLM into six categories.
\begin{description}
    \item[Wrong Character Recognition] Misrecognition of text, such as misreading signs.
    \item[Wrong Object Recognition] Misidentification of objects in the scene.
    \item[Nonexistent Objects and Texts] Mistakenly recognizing objects or text that are not present. Note that this differs from the previous two categories, where some similar objects or text were actually present.
    \item[Misunderstanding User Input] Misinterpreting a user’s question in conversation mode, such as providing an environmental description when asked to read text from a panel.
    \item[Inaccurate User Input] Errors made when the user asked about objects or text that were not present.
    \item[No Error] Accurate responses with no errors.
\end{description}
\red{
When multiple errors occur in a single sentence, errors of the same type are grouped together and counted as one. 
Errors of different types are counted separately. 
For instance, if there are multiple text recognition errors in a single sentence, they are counted as one text recognition error. 
If a sentence contains both text recognition errors and object recognition errors, each is counted separately as one text recognition error and one object recognition error.
Thus, note that the total number of errors may not match the total number of outputs.}

The results showed that 28.6\% of the outputs contained some form of error during scene descriptions whereas 60.3\% of conversation mode outputs had errors.
This difference is likely because users in conversation mode often asked for more detailed explanations, which led MLLM to attempt more complex responses and, as a result, made more mistakes.
This was particularly evident in the \textit{Nonexistent Objects and Texts} category, which accounted for only 0.07\% of errors during scene descriptions but significantly higher at 28.3\% in conversation mode.
This means that MLLM often generated descriptions of objects or text that did not exist in the environment when asked for more detailed information.
Character recognition errors were common in both modes, likely due to MLLM’s limitation in reading distant text. 
In a general sense, instead of complete failures, MLLM often partially misread the text or misidentified objects with similar-looking ones (\eg, mistaking a tall table for a reception desk).
Nevertheless, over 70\% of responses in the auto mode were accurate, demonstrating the overall usefulness of the system.

\subsection{Analysis of Usage of Each Description Level}
\label{sec:description_level_analysis}
In Tab.~\ref{tab:description_level}, we report the statistics of how much time participants spend their time using each description level.
The result shows that there were three types of usage during the study. 
P15 only used Balanced-Length mode, P11 and P14 used Balanced-Length mode most of the time while sometimes using Detailed mode, and P12 and P13 used Detailed mode most of the time. 

\begin{figure*}
    \centering
    \includegraphics[width=0.8\linewidth]{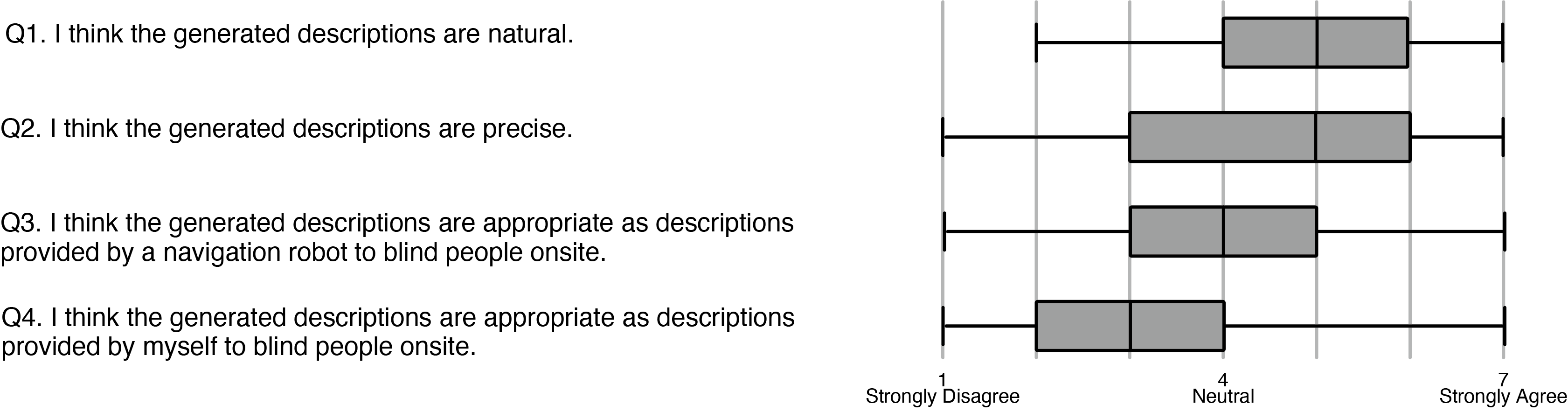}
    \caption{Box plot of evaluation with human experts in seven-point Likert points.}
    \Description{The figure displays box plots for four questions evaluating the generated descriptions: Q1. I think the generated descriptions are natural: The median score is 5, with a minimum of 2, a first quartile of 4, a third quartile of 6, and a maximum of 7. Q2. I think the generated descriptions are precise: The median score is 5, with a minimum of 1, a first quartile of 3, a third quartile of 6, and a maximum of 7. Q3. I think the generated descriptions are appropriate as descriptions provided by a navigation robot to blind people onsite: The median score is 4, with a minimum of 2, a first quartile of 3, a third quartile of 5, and a maximum of 7. Q4. I think the generated descriptions are appropriate as descriptions provided when I am there to explain to blind people onsite: The median score is 3, with a minimum of 1, a first quartile of 2, a third quartile of 4, and a maximum of 7.    
    }
    \label{fig:boxplot}
\end{figure*}

\begin{table*}[]
\caption{Rating to seven-point Likert score questions (1: strongly disagree; 4: neutral; 7: strongly agree).}
\Description{The table presents ratings from five participants (P11, P12, P13, P14, and P15) on a set of questions based on a seven-point Likert scale, where 1 indicates "strongly disagree," 4 is "neutral," and 7 is "strongly agree." The median rating for each question is also provided. The following summarizes the responses for each question. For Q1, "I was able to explore the facility," P11 and P13 rated 4, while P12, P14, and P15 rated 6, resulting in a median score of 6. For Q2, "I was able to enjoy the exploration," P11 and P13 rated 4, while P12, P14, and P15 rated 6 or 7, with a median score of 6. For Q3, "I was able to gain an interest in the things around me," P11 rated 4, P13 rated 6, and P12, P14, and P15 rated 6, leading to a median score of 6. For Q4, "The interface of the system was easy to understand," P11, P14, and P15 rated 5, while P12 and P13 rated 6, resulting in a median score of 5. For Q5, "I want to explore where I am familiar with this system," P11, P12, and P14 rated 7, while P13 and P15 rated 6, leading to a median score of 7. For Q6, "I want to explore where I am unfamiliar with this system," P11, P14, and P15 rated 7 or 6, resulting in a median score of 6.}
\label{tab:likert}

\begin{tabular}{l|ccccc|c}
\toprule
\multicolumn{1}{c|}{}                                         & P11 & P12 & P13 & P14 & P15 & Median \\ \hline
Q1. I was able to explore the facility.                       & 4   & 6   & 4   & 6   & 6   & 6      \\
Q2. I was able to enjoy the exploration.                      & 4   & 7   & 4   & 6   & 6   & 6      \\
Q3. I was able to gain an interest in the things around me.   & 4   & 6   & 6   & 6   & 6   & 6      \\
Q4. The interface of the system was easy to understand.       & 5   & 6   & 6   & 5   & 5   & 5      \\
Q5. I want to explore where I am familiar with this system.   & 7   & 7   & 6   & 7   & 6   & 7      \\
Q6. I want to explore where I am unfamiliar with this system. & 7   & 6   & 6   & 7   & 6   & 6      \\ \bottomrule
\end{tabular}%
\end{table*}
\begin{table}[]
\caption{Scores for the Raw TLX provided by each participant. Lower total scores indicate a lower workload. Each item is scored on a scale from 1 to 10, where 1 represents a lower level, and 10 represents a higher level of Mental Demand, Physical Demand, Temporal Demand, Effort, and Frustration. For Performance, 1 indicates good performance, and 10 indicates poor performance.}
\Description{
The table presents Raw TLX scores from five participants (P11, P12, P13, P14, and P15), measuring subjective workload across six dimensions: Mental Demand, Physical Demand, Temporal Demand, Performance, Effort, and Frustration. Each dimension is rated on a scale from 1 to 10, where higher scores generally indicate higher levels of demand, effort, or frustration, except for Performance, where a higher score indicates poorer performance (1: Good, 10: Poor). Lower total scores signify a lower overall workload.
Participant P11 reported low levels of Mental Demand (2), Physical Demand (2), Temporal Demand (2), and Effort (3). However, they scored higher in Performance (7) and Frustration (8), leading to a total score of 24. Participant P12 had low scores across all dimensions: Mental Demand (2), Physical Demand (2), Temporal Demand (3), Performance (2), Effort (2), and Frustration (4), resulting in the lowest total score of 15. Participant P13 scored higher in Mental Demand (5), Effort (5), and Performance (5), moderate in Physical Demand (3), and low in Temporal Demand (1) and Frustration (1), totaling a score of 20. Participant P14 reported low Mental Demand (3) and Physical Demand (2), but higher Temporal Demand (5), Performance (7), Effort (6), and Frustration (3), culminating in a total score of 26. Participant P15 had low Mental Demand (2) and Temporal Demand (2), but higher Physical Demand (6), Performance (5), Effort (6), and Frustration (7), leading to the highest total score of 28.
}
\label{tab:tlx}
\begin{tabular}{l|ccccc|c}
\toprule
                & P11 & P12 & P13 & P14 & P15 & Median \\ \hline
Mental Demand   & 2   & 2   & 5   & 3   & 2   & 2      \\
Physical Demand & 2   & 2   & 3   & 2   & 6   & 2      \\
Temporal Demand & 2   & 3   & 1   & 5   & 2   & 2      \\
Performance     & 7   & 2   & 5   & 7   & 5   & 5      \\
Effort          & 3   & 2   & 5   & 6   & 6   & 5      \\
Frustration     & 8   & 4   & 1   & 3   & 7   & 4      \\ \hline
Total Score     & 24  & 15  & 20  & 26  & 28  &        \\ \bottomrule
\end{tabular}
\end{table}

\subsection{Scene Description Quality Evaluation}
\label{sec:quality_eval}
Finally, to analyze the quality of the MLLM-generated scene descriptions from the human expert perspective, we conducted a survey with human museum guides and asked them to evaluate using a seven-point Likert scale. 
The participants were presented with images captured by the robot, each accompanied by its corresponding generated description, and were asked to evaluate the descriptions in a survey, as shown in Fig.~\ref{fig:boxplot}. 
The survey was conducted in a counterbalanced manner to mitigate potential biases.
During the main study, 164 descriptions were \rrred{generated}, and we randomly sampled half (82) of the total descriptions for evaluation.
\rrred{
The randomly sampled descriptions contain mixed levels of detail.
}
Each description is evaluated by three to four participants.
In total, 56 museum guides participated in the evaluation, with each randomly assessing five descriptions. 
There were 32 males and 20 females, and four participants did not report their gender.
Their average age was 39.6 years, with an average of 5.9 years of experience as a museum guide. 
On seven-point Likert scale items, the median self-reported familiarity with museums was 5.0, and the familiarity with LLMs was 4.0 (1: very unfamiliar, 4: neutral, and 7: very familiar).
Our analysis revealed that the experts generally perceived the generated descriptions as somewhat natural (Q1) and precise in describing an image (Q2) as shown by their median of five.
Meanwhile, they found the generated descriptions less suitable as image descriptions for blind people (Q3) and as onsite descriptions provided by experts for blind people (Q4).

\subsection{Usability and Workload Evaluation}
In Tab.~\ref{tab:likert}, we report the results of seven-point Likert items. 
For Likert items, a median score of five or higher indicates that participants generally responded positively.
The total SUS for P11 to P15 were 72.5, 80, 90, 82.5, and 77.5, respectively, showing acceptable usability of all being above 70~\cite{bangor2009determining}. 
The total Raw TLX scores for P11 to P15 were 24, 15, 20, 26, and 28, respectively. 
We show the distribution of Raw TLX scores in Tab.~\ref{tab:tlx}.
Raw TLX~\cite{byers1989traditional}, a simplified version of NASA TLX~\cite{hart2006nasa}, is known to have a high correlation with NASA TLX, and the total NASA-TLX scores for people with special needs typically ranged from 26 to 48 in previous research~\cite{hertzum2021reference}. 
Overall, our total Raw TLX scores may suggest that participants did not experience a significant load during the task.
We also observed that the median value for mental, physical, and temporal demand was relatively lower, scoring 2. 
This is likely due to the robot navigating them, allowing participants to explore without being burdened by these demands. 
Nonetheless, a relatively higher median value was observed for Performance, Effort, and Frustration, indicating that some users experienced a lack of satisfaction with the exploration experience provided by the system. 

\subsection{Qualitative Analysis}
\subsubsection{Positive Feedback}
All participants expressed their appreciation for the experience of wandering around a building to explore without specific destinations in mind with the help of our system:
\newanswer[\label{P12IWantThis}]\textit{``
When the camera explains things it recognizes, like how bright the room is or what the floor looks like, or what objects are placed where, I found myself nodding in agreement multiple times, like, ``Oh, so this is how it looks.'' 
I remember when I first held the suitcase robot, I deeply empathized with guide dog users. I thought, ``Oh, so this is what it's like to have a guide dog.'' However, since I can't take care of a guide dog, I’ve given up on that option. 
And now, with this navigation system that explains various situations, it's exactly what I need. It’s not just about setting a destination and getting there but feeling the freedom to explore spontaneously. For example, the ability to roam a large shopping mall freely and explore on a whim feels like true freedom to me. Instead of pre-planning every move or relying on a guide, I could simply grab my suitcase and decide to venture out spontaneously.''} (P12)

\red{
The same participant, P12, who had been to the facility previously, noted that they still had new discoveries with the system:
\newanswer[\label{P12IHaveBeen}]\textit{``
I've been to this museum before, but when the guide explained things to me back then, it was more like a general explanation about the atmosphere and such. 
But earlier with the system, there was a very detailed explanation that came out of the suitcase. Like, about how bright sunlight comes [...] 
There were things I didn’t know that made me learn new stuff, even though I thought I knew about the facility.''} (P12)
Also, P12 and P14 noted the feeling of relief not relying on sighted assistance:
\newanswer[\label{P14ThereHaveBeenNoSystem}]\textit{``
I don’t think there has ever been a system that explains your surroundings while walking. [...]
When walking with other people, I often find myself feeling a sense of obligation. I worry that they’re putting in extra effort to describe things because I can’t see. And then I feel like I have to respond to them since they’re trying so hard—which can be exhausting. But with this system, I feel I can go strolling by myself.''} (P14)
}

Participants also noted the functionality to go to an aforementioned destination and Q\&A functionality particularly useful:
\newanswer[\label{P11TakeMeBack}]\textit{``(The ``Take-Me-There'' functionality is) I think it's wonderful. After all, spatial awareness is difficult, so going back to landmarks is very important. If it is accurate, I think it's great because it can be extremely helpful for spatial cognition.''} (P11) and
\newanswer[\label{P14Q&A}]\textit{``When engaging in a conversation, not knowing what kind of response you'll get, the feeling of unease and excitement that's both a plus and a minus, I think. But I found it really great that you can still ask questions. So even if the response you get doesn't answer your question, or even if it's just ``I don't know,'' the fact that you can at least ask is important.''} (P14) 

\subsubsection{Adjusting Detail of Description}
When we discussed their preference in the level of detail of descriptions, all participants described that it would rather depend on the scenario they are in:
\newanswer[\label{P12NormalMode}]\textit{``It might depend on the location, but I know I can get detailed information in Q\&A functionality. So, for familiar places, the Balanced-Length mode might be fine. However, there are parts where I'd want the Detailed Description mode for unfamiliar places. For example, switching between modes could be useful, like having Detailed Description mode first for explanations about the room's brightness and how easy it is to walk around. ''} (P12)

\subsubsection{Comments to Improve the System}
\label{sec:improve}
Participants suggested various improvements to the system.
One particular suggestion was to incorporate functionality for the robot to understand sounds. 
As the experiment location was a science museum, various exhibits emitted sounds.
P13 noted that they would like to inquire about the sound sources, which were not supported by the system:
\newanswer[\label{P13Sound}]\textit{``We are extremely sensitive to sounds, and it becomes a point of interest. At a place like the exhibition hall we're visiting this time, various sounds are coming from all directions. This prompts questions like, ''What's happening at that sound over there?'' Therefore, it would be advantageous if we could ask specific questions like, ``What's that sound coming from the right?'' ''} (P13)

Also, four participants (P11-P13 and P15) found the descriptions from the system still insufficient to explore, as described in the following comments:
\newanswer[\label{P15Insufficient}]\textit{``The place we did the task this time was quite out of the ordinary. Even if you were walking around with my family, I think they would also have difficulty explaining it. Therefore, I felt it might still be somewhat challenging for machines to handle this kind of thing. However, I did feel it was good that I got a sense of what was there. But when it comes to the actual detailed explanations, it was not there [...]''} (P15)

\subsubsection{Specification of Proceeding Direction}
While we introduced all functionality to participants within the training session, we observed that only P15 used the functionality to specify which way to proceed via a button or conversation.
P15 tended to use the functionality when P15 was interested in a specific object:
\newanswer[\label{P15DirectionSpecification}]\textit{``It seems that when I was told, ``There's something on the right,'' I tried to approach toward it because I wanted to get closer when I used something like that.''} (P15)
\section{Discussion}
\subsection{Experience of Using WanderGuide}
WanderGuide provided participants with the experience of exploring unfamiliar indoor environments without a specific destination in their minds, mimicking the spontaneous wandering experience of sighted people (C\ref{P12IWantThis}). 
Participants expressed a sense of confidence when using the system, noting that it allowed them to navigate independently without relying on traditional tools like white canes (C\ref{P12IWantThis}). 
As described by C\ref{P15Insufficient} and the ratings of 4 from P11 and P13 to Q1 and Q2 in Tab.~\ref{tab:likert}, there still exists the limitation of being unable to describe specific information.
Thus, there is a need for further research on how to appropriately convey surrounding information to blind people.
Still, the system’s ability to deliver real-time descriptions of objects, walls, and spatial layouts enabled participants to form an imagination (C\ref{P12IWantThis}) of their surroundings, sparking their desire to use the system in familiar and unfamiliar environments (Tab.~\ref{tab:likert} Q5 and Q6).
In short, WanderGuide has the potential to provide users with an experience similar to that of navigating with sighted assistants to explore the environment, but the users can explore independently.
We believe this research opens a new frontier to the concept of \textit{map-less exploration} guide system for blind people.

\red{
\subsection{Scene Description by MLLM}
\label{sec:scene_description}
Our survey in Sec.~\ref{sec:quality_eval} revealed that descriptions by MLLM were rated high for their naturalness and suitability for general image description but were not for actual descriptions to be provided to blind people by sighted experts.
This may be because the style and content of the generated descriptions differ from those typically provided to blind people during live interactions. 
For example, museum guides often focus on explaining notable objects or visible exhibits, complementing their descriptions with additional knowledge about the exhibits.
In contrast, the generated description often lacked concrete explanation about exhibits and shops, such as their names (C\ref{P01NeedMoreConcreteInformation}, C\ref{P09NegativeImpression}, and C\ref{P15Insufficient}).
This problem may be more prominent because the study was conducted in a science museum, where each exhibit contains detailed information that is not visually apparent but needs to be explained. 
On the other hand, from participant feedback, participants noted that MLLM-generated descriptions are comprehensive (C\ref{P07Enjoy}, C\ref{P10VariousAndDifferentInformation}, C\ref{P12IHaveBeen}, and C\ref{P14ThereHaveBeenNoSystem}), and provide them with enjoyment (C\ref{P07Enjoy}) and imagination of vision perception (C\ref{P10VariousAndDifferentInformation}).
They noted that MLLM provided them with information that they usually do not get from sighted assistants, leading to new discoveries (C\ref{P12IHaveBeen}). The descriptions provided by MLLM additionally allow blind people to tune in without hesitation and the need to rely on sighted people (C\ref{P14ThereHaveBeenNoSystem}).
These results indicate that evaluation from sighted experts may be stricter than that from blind people.
Nonetheless, these results suggest that MLLM for blind people's exploration could be further enhanced by providing more specific information about surrounding shops or exhibits, potentially inferring details when necessary. 
}

\subsection{Personal Preferences}
The studies revealed distinct preferences among participants regarding the levels of detail in the descriptions (Sec.~\ref{sec:Implementation_description_mode}) and interaction modes (Sec.~\ref{sec:implementation_button}). 
From the formative study, participants were divided into three preference groups, highlighting users' diverse information needs regarding exploration and goal-oriented navigation (Sec.~\ref{sec:implication_varydetail}).
Differences in preferences were mainly attributed to personality traits, because participants who were ``Destination-Oriented'' (Tab.~\ref{tab:demographics1}), or were mostly concerned with reaching destinations, mentioned they did not enjoy the detailed explanation of the system and preferred short, concise information. 
For example, one early blinded participant mentioned that exploration did not interest him, as he had barely done it in his daily life (C\ref{P02Conditioned}). 
On the other hand, some participants enjoyed imagining the scenes conveyed by the system. 
Congenital users commented that the descriptions felt as if they were actually seeing the surroundings, while acquired users likened it to their recalled experiences when they could still see. 
Interestingly, those who particularly enjoyed the system and were ``Exploration-Inclined'' were all female, while the Intermediate group, who enjoyed exploration but wanted more control over the information provided, consisted mainly of male participants.
We note that ``Destination-Oriented'' users expressed dissatisfaction with the system because they felt the scene description capabilities of the MLLM did not meet their expectations for exploration. Therefore, if the system was improved and was able to convey more concrete information, they might express different opinions. 

In the main study, further differences regarding how users interacted with the system were observed. 
Firstly, we observed that participants adjusted the system's levels of description, demonstrating our design aligns with their needs, which were based on three types of preferences identified in the formative study (Sec~\ref{sec:description_level_analysis}). 
The variation in the portions used for each mode further underscores the need for configurable descriptions. 
Also, how they used the conversation mode varied. 
Three participants frequently asked questions to the system to gather information about their surroundings (Sec.~\ref{sec:activity_breakdown}), while P15 preferred having more manual control over the robot’s navigation. 
Meanwhile, P12 favored the auto mode, where the robot guided them with minimal intervention. 
These observations highlight the need to consider customizing to various dimensions of personal preference, from description details to user autonomy, for future development.

\subsection{Design Implications and Future Development Directions} 
\label{sec:design_implication}
Two key design implications were observed in our studies. 
First, allowing the users to control the level of detail in the scene descriptions emerged as one of the most important design requirements. 
The system may benefit from further \textit{personalization} by users verbally describing their personal information needs as in previous research~\cite{Kaniwa2024ChitChatGuide}. 
Second, participants expressed the need for audio-based recognition capabilities, especially in environments where sound is an integral part of the experience, such as museums (C\ref{P13Sound}). 
The ability to answer questions about sounds and potentially guide users to the sounds' sources would enhance their exploration experience.

On the development side, the primary challenge encountered throughout the two studies was the system’s inability to provide detailed information that participants required, particularly regarding the identification of POI-related objects, as described in \red{Sec.~\ref{sec:scene_description}.}
We attempted to address this by upgrading the robot's hardware, \ie, adding a 1080p resolution fisheye camera to a much higher position. 
Still, participants found the descriptions lacking in detail and conveyed information somewhat vague, as partially shown by the ratings of 4 from P11 and P13 to Q1 Tab.~\ref{tab:likert}. 
We deduce that this was because the captured images sometimes did not contain useful information, such as the names of certain objects, or because the MLLM failed to accurately identify the useful information.
As a possible improvement, the robot could utilize history images by selecting the image with the best view to generate descriptions. 
Also, the robot could utilize, other modalities, such as colored point clouds by fusing camera images with the LiDAR sensor to provide three-dimentional sensor details to MLLM~\cite{liu2024uni3d,xu2023pointllm}. 
In conclusion, the MLLM module is the bottleneck of our system's technological development.
Similar system development efforts in the future should allocate the most resources to tackling this technological challenge.
Still, the issue may be gradually solved as MLLM is the current core area actively developed by researchers.

Another significant challenge in development we encountered is the challenge of running map-less navigation algorithms in diverse novel environments, which requires extensive development. 
Incorporating vision modalities~\cite{chang2024goat}, which we did not use in this study, could potentially enhance the robot's navigation capabilities. 
Achieving this, however, demands human-level object and layout recognition and real-time processing speed, where further research is required.

\subsection{Limitation and Future Work}
We were unable to examine user preferences over the long term, as participants in our study interacted with the system only for a short duration (20-40 minutes in the formative study and 70 minutes in the main study). 
Only a small portion of the reliance on concise descriptions may be due to the study's design limiting participants' time to explore.
The time constraints may have led users to act on the cost-effective information acquisition.
However, if the system is used regularly, users may encounter more situations where they prefer to use the concise mode, as indicated by C\ref{P12NormalMode}. 
Also, their preferences might change as they become more adept at utilizing it as a tool to query information, which the MLLM is particularly proficient at.
Thus, future research should investigate the effects of long-term use of the system.

We conducted two studies in two indoor locations. 
To capture more diverse needs, future studies should also explore the system’s performance in more diverse environments. 
This may reveal various additional information needs. 
The usage of the wheeled robot, while beneficial in guiding blind users because it is silent~\cite {wang2022can}, remains a constraint when navigating stairs or uneven terrain. 
This limitation, however, could be alleviated through user collaboration, such as assisting the robot in getting onto elevators or slightly lifting the robot over small steps.
Thus, future research should investigate the system devices's capabilities in different environments, as well as how these robots can address physical limitations by interacting with users.
\red{
Finally, for the main study, we were unable to conduct it in crowd environments with bystanders potentially obstructing the cameras, because the primary study was conducted in the science museum outside of regular operational hours. 
Handling crowded environments with robots, even when prebuilt maps are used, remains a significant challenge in the field of robotics~\cite{wang2022group}. 
Therefore, in future work, we aim to address the usability limitations of our system in such scenarios by integrating novel algorithms designed to manage crowded environments~\cite{wang2022group}.
}

The MLLM often made mistakes or referred to non-existent objects, with these errors being particularly noticeable in its responses within the Q\&A functionality (Sec.~\ref{sec:error_analysis}). 
The most common misrecognitions involved either partially reading the text or confusing objects with similar-looking ones. 
However, the performance of the MLLM is not the primary focus of our research. 
To ensure users receive the most accurate information possible, we will continue updating the MLLM used in the system.
\red{Also, some of the image inputs provided to the MLLM may have been affected by motion blur, potentially leading to a degradation in the quality of the generated descriptions. 
This issue could be addressed by using cameras that are more resistant to motion blur or by implementing algorithms that detect motion blur and select alternative frames for processing.}

Recruitment was conducted through our institution's email list, which includes many participants from previous studies. 
We acknowledge that these participants may have exhibited a positive bias toward our study, as they had expectations regarding the development of the robot system.
\red{
Furthermore, we obtained valuable insights from five participants, and involving more participants might have provided additional perspectives. 
Given the difficulty of recruiting many blind participants, we chose to iterate the study with five participants in each study, rather than conducting a single study with a larger group.
}

\section{Conclusion}
Towards realizing a scalable map-less guide system that assists blind people in exploring, we developed WanderGuide, a robotic guide  system designed to provide real-time descriptions of surroundings and to offer conversation functionalities that allow users to specify their destinations or ask questions.
The formative study with ten blind participants revealed that there are three types of preferences over the levels of details of the descriptions generated by the system.
In a subsequent main study with five blind participants, all of them expressed appreciation for the experience of wandering freely without a fixed destination, as well as a desire to use the system for exploring both familiar and unfamiliar areas. 
The study further revealed that including audio recognition would be the immediate next step for developing our system. 
It also revealed that customizing to diverse user preferences is important and that MLLM is the key bottleneck of the technology development of our system.
We hope this research contributes to the potential deployment of robotic guide systems in general use cases, enabling blind users to explore independently.

\begin{acks}
We would like to thank all the participants in our user study.
We are also deeply thankful to Mori Building Co., Ltd. for providing the experimental location.
Finally, we thank all members of Miraikan, including Hironobu Takagi and Hiromi Kurokawa,  and
the Consortium for Advanced Assistive Mobility Platform for their support.
This work was supported by JSPS KAKENHI (JP23KJ2048).
\end{acks}

\bibliographystyle{ACM-Reference-Format}
\bibliography{main}

\appendix

\lstset{
  backgroundcolor=\color{gray!20}, 
  basicstyle=\ttfamily\footnotesize, 
  breaklines=true,                 
  frame=single,                    
  framerule=0pt,                   
  xleftmargin=5pt, xrightmargin=5pt 
}

\section{Appendix: Prompts to MLLM}
In this section, we list full prompts to MLLM and LLM, which were used in this paper.

\subsection{Prompt Used For Translating Native Language to English}
\label{appendix:translate}
As the research was conducted in a country where English is not spoken, we used the below prompt to translate any data obtained in the native language throughout the paper. 
Note that the authors manually refined the output to keep the nuances of the original language.
This prompt was also used to translate the prompt engineered in the native language, which was fed into the MLLM for generating scene descriptions.
\begin{lstlisting}
Please translate the given <Native Language> to English. Make sure to keep the nuances and context of the original text.
<Native Language>:  Text written in native Language
English:
\end{lstlisting}

\subsection{Prompt Used In The Formative Study}
\label{appendix:prompt_formative}
Below is the prompt used to generate descriptions in the formative study.
\begin{lstlisting}
# Instructions  
Please describe the image.  
The text you generate will be read directly to visually impaired individuals. Make sure your description is engaging so that visually impaired individuals can enjoy listening to it.  
To describe the image, you must follow the rules outlined below.  

## Rules you must strictly follow to comply with the instructions  

### Rules on what you should do  
1. Since visually impaired individuals will listen while walking, provide a description in one cohesive sentence. Please describe as many objects and their details as possible.  
2. Generate the description in 1 to 4 sentences in total.  
3. If necessary, first describe the overall layout or the general view of the location.  
4. After that, identify the objects located on the left, in front, and on the right of the image, and explain the information required to understand the scene.  
5. Always describe the scene in the following order: overall view, left side, front, right side.  
6. When describing, use a tone similar to a guide for the visually impaired, such as "On the right, there is..."  
7. If there is a store, make sure to include information about what the store offers (for example, the type of cuisine if it is a restaurant). Also, include a description of the store's atmosphere (e.g., bright, calm).  
8. Only describe objects that are clearly visible. Include descriptions of distinctive objects.  
9. Create a description that is enjoyable to listen to and allows the listener to learn about their surroundings.  

### Rules on what you should not do  
10. Avoid unnatural words for the listener, such as "the image" or "viewpoint."  
11. You do not need to include common and unremarkable objects (e.g., tables and chairs in a restaurant) in the description.  
12. If there is nothing to describe in a particular direction (e.g., there is nothing on the right), you do not need to mention that direction.  
13. Do not describe the floor, ceiling, shadows, distant unclear objects, or the brightness or darkness of the lighting.  

## Response Format  
If you generate a good description that follows the above rules, you will receive a tip.  
Please respond in JSON format.  
Include the image description under the "description" key.  
Start your response with ```json\n{ to indicate the beginning of the JSON.  

Here is an example of a response:  
```json  
{  
"description": "<description of the surroundings>",  
}  
```
\end{lstlisting}

\subsection{Prompts Used In The Formative Study}
\label{appendix:prompt_main}
This section provides prompts used in the main study.

\subsubsection{The Prompt for Generating Detailed Description}
Below is the prompt used to generate a detailed description.
\begin{lstlisting}
# Instructions  
Please describe the image.  
You are given three images that provide a view of your left, right, and front, as well as a view from a fisheye camera that captures the overall view from a high point of view.
The text you generate will be read directly to visually impaired individuals.  
When writing the description, please aim to make it appealing so that it creates an enjoyable experience for the listener.  
The most important thing is to provide detailed and specific information so that the listener can feel as if they are actually at the scene.  
Being specific means describing the category or name of objects, their condition, and the role they play.  
For example, a description like "circular wooden object" is not specific, but "a circular wooden table with YYY written on the nearby guide" is specific.  
Similarly, "iron exhibit" is vague, while "a tall, iron exhibit, possibly XXX" is specific.  
When describing the image, you must follow the rules below.

## Rules that must be followed to comply with the instructions  
1. The description must be something that a visually impaired person can listen to while walking. Provide a coherent description in one block of text. Explain as many objects and their details as possible.  
2. Keep the description to 3-4 sentences at most (120-240 characters).  
3. Use polite language (honorifics).  
4. Identify and describe objects located in the overall scene, to the left, front, and right that are necessary to understand the scene.  
5. Only describe clearly visible objects. Include distinctive objects in your description.  
6. Always describe the overall scene first, followed by objects on the left, in front, and then on the right.  
7. Strive to include the following information:
    - Details about the building's interior and decoration.
    - Information about the layout of the building (such as whether the front is open, where walls are, and the directions one can go).
    - Information on the surrounding brightness and the amount of light coming through windows.
    - Information about people in the surroundings, their actions, clothing colors, and whether they are staff or customers.
    - If it's a store, provide information on whether the entrance is open like a terrace and whether guide dogs can wait there.
    - Include information on visible stores or exhibits. Be sure to mention their category (e.g., the type of food if it's a restaurant, or what kind of place the exhibit is). If possible, include the name of the place. For exhibits, state whether they are interactive or for viewing only.
    - When describing objects, be specific (mention the category or name). For example, if there's a counter, specify if it looks like a cafe counter.
    - Mention people walking toward the front if there's a risk of collision.
    - Use numbers when explaining object positions (e.g., "5 meters to the right").
    - If there is a sign or guidepost, describe what it is and read out the text written on it.
    - Read out visible text.
    - Use adjectives like futuristic, stylish, modern, or classic to make the exploration more enjoyable and to help the listener visualize the scene.
8. Do not use unnatural words for the listener like "image," "viewpoint," or "overall."
9. If there's nothing to describe in a certain direction (e.g., nothing on the right side), do not describe that direction.  
10. Do not summarize or conclude with a description of the overall direction or scene when finishing the explanation.  
11. Do not describe anything not visible in the image. Do not lie or hallucinate details.

## Response Format  
If you follow the rules above, you will receive a tip.  
If you ignore the rules, you will be penalized and have to pay a fine.  
Please do your best to comply with these instructions.

Respond in JSON format.  
First, include the initial description of the image under the "initial_description" key.  
Next, include points for improvement under the "improve_thoughts" key.  
Finally, include the revised image description under the "description" key.  
Start your response with `json\n{`.  
Here is an example response:

```json
{
"initial_description": "<initial description>",
"improve_thoughts": "<points for improvement>",
"description": "<revised description>",
}
```

\end{lstlisting}

\subsubsection{The Prompt for Generating Balanced-Length Description}
Below is the prompt used to generate a balanced-length description.
\begin{lstlisting}
# Instructions  
Please describe the image.  
You are given three images that provide a view of your left, right, and front, as well as a view from a fisheye camera that captures the overall view from a high point of view.
The text you generate will be read directly to visually impaired individuals.  
Keep the description concise, but aim to make it appealing and enjoyable for the listener.  
The most important thing is to provide detailed and specific information so that the listener can feel as if they are actually at the scene.  
Being specific means describing the category or name of objects, their condition, and the role they play.  
For example, a description like "circular wooden object" is not specific, but "a circular wooden table with YYY written on the nearby guide" is specific.  
Similarly, "iron exhibit" is vague, while "a tall, iron exhibit, possibly XXX" is specific.  
When describing the image, you must follow the rules below.

## Rules that must be followed to comply with the instructions  
1. The description must be something that a visually impaired person can listen to while walking. Provide a coherent description in one block of text. Explain as many objects and their details as possible.  
2. Keep the description to 2-3 sentences at most (60-120 characters).  
3. Use polite language (honorifics).  
4. Identify and describe objects located to the left, front, and right that are necessary to understand the scene.  
5. Only describe clearly visible objects. Include distinctive objects in your description.  
6. Always describe objects in the following order: left, front, and right.  
7. Strive to include the following information:
    - If it's a store, provide information on whether the entrance is open like a terrace and whether guide dogs can wait there.
    - Include information on visible stores or exhibits. Be sure to mention their category (e.g., the type of food if it's a restaurant, or what kind of place the exhibit is). If possible, include the name of the place. For exhibits, state whether they are interactive or for viewing only.
    - When describing objects, be specific (mention the category or name). For example, if there's a counter, specify if it looks like a cafe counter.
    - Mention people walking toward the front if there's a risk of collision.
    - Use numbers when explaining object positions (e.g., "5 meters to the right...").
    - If there is a sign or guidepost, describe what it is and read out the text written on it.
    - Read out visible text.
8. Do not use unnatural words for the listener like "image," "viewpoint," or "overall."
9. If there's nothing to describe in a certain direction (e.g., nothing on the right side), do not describe that direction.  
10. Do not summarize or conclude with a description of the overall direction or scene when finishing the explanation.  
11. Do not describe objects if you cannot provide specific information about them.  
12. Do not include information about people in the surroundings unless there is a risk of collision.  
13. Do not include information about the amount of light or brightness in the surroundings.  
15. Do not use subjective adjectives like futuristic, stylish, modern, or classic.  
16. Do not describe anything not visible in the image. Do not lie or hallucinate details.

## Response Format  
If you follow the rules above, you will receive a tip.  
If you ignore the rules, you will be penalized and have to pay a fine.  
Please do your best to comply with these instructions.

Respond in JSON format.  
First, include the initial description of the image under the "initial_description" key.  
Next, include points for improvement under the "improve_thoughts" key.  
Finally, include the revised image description under the "description" key.  
Start your response with `json\n{`.  
Here is an example response:

```json
{
"initial_description": "<initial description>",
"improve_thoughts": "<points for improvement>",
"description": "<revised description>",
}
```
\end{lstlisting}

\subsubsection{The Prompt for Generating Concise Description}
Below is the prompt used to generate a concise description.
\begin{lstlisting}
# Instructions  
Please describe the image.  
You are given three images that provide a view of your left, right, and front, as well as a view from a fisheye camera that captures the overall view from a high point of view.
The text you generate will be read directly to visually impaired individuals.  
The description should be concise and minimal, allowing the listener to quickly understand their surroundings.  
Visually impaired individuals are listening to the image description to locate their destination.  
The most important thing is to provide detailed and specific information so that the listener can feel as if they are actually at the scene.  
Being specific means describing the category or name of objects, their condition, and the role they play.  
For example, a description like "circular wooden object" is not specific, but "a circular wooden table with YYY written on the nearby guide" is specific.  
Similarly, "iron exhibit" is vague, while "a tall, iron exhibit, possibly XXX" is specific.  
When describing the image, you must follow the rules below.

## Rules that must be followed to comply with the instructions  
1. The description must be something that a visually impaired person can listen to while walking. Provide a coherent description in one block of text.  
2. Keep the description to 1-2 sentences at most (0-60 characters).  
3. Use polite language (honorifics).  
4. Identify and describe objects located to the left, front, and right that are necessary to understand the scene.  
5. Only describe clearly visible objects. Include distinctive objects in your description.  
6. Always describe objects in the following order: left, front, and right.  
7. Strive to include the following information:
    - If it's a store, provide information on whether the entrance is open like a terrace and whether guide dogs can wait there.
    - Include information on visible stores or exhibits. Be sure to mention their category (e.g., the type of food if it's a restaurant, or what kind of place the exhibit is). If possible, include the name of the place. For exhibits, state whether they are interactive or for viewing only.
    - Use numbers when explaining object positions (e.g., "5 meters to the right...").
    - If there is a sign or guidepost, describe what it is and read out the text written on it.
    - Read out visible text.
8. Only convey specific information.
9. Keep the description short, direct, and concise.  
10. Do not use unnatural words for the listener like "image," "viewpoint," or "overall."
11. If there's nothing to describe in a certain direction (e.g., nothing on the right side), do not describe that direction.  
12. Do not summarize or conclude with a description of the overall direction or scene at the beginning or end of the explanation.  
13. Do not describe decorations. Simply convey what is there and provide specific information.  
14. To keep the length minimal, do not include subjective adjectives.  
15. Do not include unnecessary information that does not help the listener locate their destination (e.g., details about furniture such as chairs or tables).  
16. Do not describe objects if you cannot provide specific information about them.  
17. Do not include information about people in the surroundings unless there is a risk of collision.  
18. Do not include information about the amount of light or brightness in the surroundings.  
19. Do not describe anything not visible in the image. Do not lie or hallucinate details.

## Response Format  
If you follow the rules above, you will receive a tip.  
If you ignore the rules, you will be penalized and have to pay a fine.  
Please do your best to comply with these instructions.

Respond in JSON format.  
First, include the initial description of the image under the "initial_description" key.  
Next, include points for improvement under the "improve_thoughts" key.  
Finally, include the revised image description under the "description" key.  
Start your response with `json\n{`.  
Here is an example response:

```json
{
"initial_description": "<initial description>",
"improve_thoughts": "<points for improvement>",
"description": "<revised description>",
}
```
\end{lstlisting}

\end{document}